\documentclass[12pt]{iopart}
\usepackage{iopams}  
\usepackage{graphics}
\usepackage{epsfig}
\usepackage{cite}
\eqnobysec

\newcommand{\mexp}[1]{e\sp{#1}}
\newcommand{\mfrac}[2]{ {#1} / {#2} }
\def\C{{\mathbb{C}}}
\def\lambdabar {\mathchar'26\mkern-10mu\lambda}
\def\binom#1#2{{#1 \choose #2}}
\def\qbinom#1#2{\left[ \matrix{#1\cr #2}\right]}

\begin{document}

\title{The Wigner function of a $q$-deformed harmonic oscillator model}

\author{E.I. Jafarov$^{1,2}$\footnote[3]{E-mail: azhep@physics.ab.az}, S. Lievens$^{2}$\footnote[4]{To whom correspondence should be addressed (E-mail: Stijn.Lievens@UGent.be)}, S.M. Nagiyev$^{1}$ and J. Van der Jeugt$^{2}$\footnote[5]{E-mail: Joris.VanderJeugt@UGent.be}}

\address{${}^{1}$\ Institute of Physics, Azerbaijan National Academy of Sciences, Javid av. 33, AZ1143, Baku, Azerbaijan}

\address{${}^{2}$\ Department of Applied Mathematics and Computer Science, Ghent University, Krijgslaan 281-S9, B-9000 Gent, Belgium}


\begin{abstract}
The phase space representation for a $q$-deformed model of the quantum harmonic oscillator is constructed. We have found explicit expressions for both the Wigner and Husimi distribution functions for the stationary states of the $q$-oscillator model under consideration. The Wigner function is expressed as a basic hypergeometric series, related to the Al-Salam-Chihara polynomials. It is shown that, in the limit case $h \to 0$ ($q \to 1$), both the Wigner and Husimi distribution functions reduce correctly to their well-known non-relativistic analogues. Surprisingly, examination of both distribution functions in the $q$-deformed model shows that, when $q \ll 1$, their behaviour in the phase space is similar to the ground state of the ordinary quantum oscillator, but with a displacement towards negative values of the momentum. We have also computed the mean values of the position and momentum using the Wigner function. Unlike the ordinary case, the mean value of the momentum is not zero and it depends on $q$ and $n$. The ground-state like behaviour of the distribution functions for excited states in the $q$-deformed model opens quite new perspectives for further experimental measurements of quantum systems in the phase space.
\end{abstract}


\submitto {\JPA}


\section{Introduction}

The harmonic oscillator concept occupies a central position in science and
engineering due to its simplicity and exact solubility in both classical and
quantum descriptions. Today it appears in mechanics, electromagnetism,
electronics, optics, acoustics, astronomy, nuclear theory, to name just a few~\cite{moshinsky}.

In the quantum approach the harmonic oscillator is one of the exactly solvable
problems studied in detail due to its considerable physical interest and 
applicability~\cite{landau}. 
To examine the applicability of a system modelled by a harmonic oscillator, one 
wishes to compare theoretical results about position and momentum with measurements.
In practical quantum applications however, one is faced with measurability problems 
of the quantum system states due to the uncertainty principle of quantum physics.
For this reason, it is common to work with area elements in phase space, 
whose size is not smaller than Planck's constant, and obtain an
expression for the joint probability distribution function of momentum and position.
Then, having a certain method for the measurement of this quantity, the gathered results will
provide a complete characterization of the quantum system under consideration, and allow
the use of more transparent classical language to examine its properties.

The Wigner distribution function is the main theoretical tool to construct the
phase space of a quantum system. It is the closest quantum mechanical analogue of
the classical probability distribution over the phase space, i.e.\ $\mathop
{\lim }\limits_{\hbar \to 0} W \left(p,x \right) = \rho \left(p,x \right)$. This distribution
function was first proposed by Wigner to study the quantum correction for a
thermodynamic equilibrium~\cite{wigner}. Quantum harmonic oscillators are one of first
systems where analytical expressions for the Wigner distribution
function were calculated, for stationary states as well as for states of the
thermodynamic equilibrium. These expressions are important due to
experimental measurements done in the past two decades~\cite{smithey,mcmahon}. 

For example, the novel experimental method of optical homodyne tomography 
allows the measurement of the Wigner function for squeezed, vacuum, one- and two-photon 
states~\cite{smithey,lvovsky,ourjoumtsev}. 
Note also experiments where measurements of the Wigner function of Fock states 
were performed on vibrational states of a trapped $Be^+$ ion~\cite{leibfried}. 
In other words, advanced measurement techniques for the Wigner function 
changed its status from an \lq\lq only theoretically computable expression\rq\rq\ to that of an 
\lq\lq also directly measurable quantity\rq\rq.

All of this, however, is still not applicable to generalizations of
the quantum harmonic oscillator, in particular the so-called $q$-deformed harmonic
oscillator. 
The origin of $q$-deformed oscillators can be traced back to the work of Iwata~\cite{iwata}, 
who generalized the Heisenberg commutation relation and solved the relevant
eigenvalue problem.  Later on, in the 70's $q$-deformed generalizations of the
harmonic oscillator were quite popular and since the end of the 80's they are
back into fashion.  
It is interesting to observe that $q$-deformed
harmonic oscillators models are also simple and exactly solvable. Moreover,
the appearance of an extra parameter $q$ gives rise to additional application
opportunities compared to the non-relativistic harmonic oscillator. 
Deformed oscillators have found applications in the theory of quons~\cite{greenberg}, 
statistical physics~\cite{chaturvedi}, generalized thermodynamics~\cite{lavagno} 
and in models describing a small violation of the Pauli exclusion principle~\cite{ignatiev}, to name a few. 
However, in spite of these applications, many aspects of the $q$-oscillator model have not been developed. 
One of the main reasons restricting its wide applications is the absence of an exact expression in terms of phase space. In this context, it is necessary to note the work in~\cite{rajagopal,zhang,galetti}, 
where the possibility of obtaining the Wigner function associated with the $q$-Heisenberg 
commutation relation is discussed. 
However, in spite of attempts to derive an analytical expression of the
Wigner function for the $q$-deformed harmonic oscillator models, up till now
this goal was not achieved. 
The main problem is related to difficulties
for the right definition of the momentum and position operators and to transformation
formulae between them in the $q$-case. 

If one imagines modern quantum physics as a puzzle, then due to the 
absence of an explicit description of $q$-deformed harmonic oscillators in
phase space, some pieces of the puzzle are missing. 
In this paper, we examine some missing pieces of this puzzle and 
present explicit expressions for the Wigner and Husimi distribution functions for the
stationary states of a $q$-deformed harmonic oscillator model. 
We also analyse these expressions for various values of the parameter $q$ 
and prove that they reduce to the well-known expressions for the non-relativistic 
harmonic oscillator in the limit $q \rightarrow 1$.

Our paper is structured as follows: in Section~2, 
we provide basic information about the definition of the Wigner function and its Gauss smoothed analogue. 
We present their explicit expressions for the stationary states of the non-relativistic linear harmonic oscillator.
Section~3 is devoted to a model of the $q$-deformed oscillator, whose wave functions are expressed in coordinate and momentum spaces by Rogers-Szeg\"o and Stieltjes-Wigert polynomials respectively. 
We present explicit expressions for the Wigner and Husimi distribution functions for the stationary states of the $q$-deformed oscillator in Section~4. Discussions and Conclusions are given in Section~5 and Section~6.

To end this introduction, we collect some common notation and known results 
for $q$-series. Many of the expressions encountered in this 
article will be expressible in terms of basic hypergeometric series ($q$-series), for
which we use the standard notation of Gasper and Rahman~\cite{GR2}. The 
$q$-shifted factorial is defined as:
\begin{equation*}
(a;q)_0 = 1,\quad\mbox{and}\quad (a;q)_n = \prod_{k=0}^{n-1} (1-a\,q^k),\quad 
\mbox{for}\quad n=1,2,\ldots,\infty,
\end{equation*}
and $(a_1,\ldots,a_m;q)_n = (a_1;q)_n\cdots(a_m;q)_n$.  A basic hypergeometric
series is defined as
$$
\fl \quad {}_r\phi _s \left( {\begin{array}{*{20}c}
   {a_1 ,a_2 , \ldots ,a_r }  \\
   {b_1 ,b_2 , \ldots ,b_s }  \\
\end{array};q,z} \right)
 =
 \sum_{k=0}^{\infty} \frac{(a_1,a_2,\ldots,a_{r};q)_k}{(q,b_1,b_2,\ldots,b_s;q)_k} \left( (-1)^k q^\binom{k}{2}
\right)^{1+s-r} z^k.
$$

In most cases, we will have that $r=s+1$.  Moreover, a basic hypergeometric
series is terminating if one of the parameters $a_j$ ($j=1,\ldots,r$) equals
$q^{-n}$ with $n$ a nonnegative integer.   The single most important 
summation formula for basic hypergeometric series is the $q$-binomial theorem:
\begin{equation}
\label{eq:qbinomial}
_1 \varphi _0 \left( {\begin{array}{*{20}c}
   a  \\
    -   \\
\end{array};q,z} \right) = \sum_{k=0}^\infty 
\frac{(a;q)_k}{(q;q)_k} z^k = \frac{ (az;q)_\infty}{ (z;q)_\infty}, \quad
|z|<1,\quad |q|<1.
\end{equation}

In the terminating case, i.e.~when $a=q^{-n}$, this reduces to
\begin{equation*}
	_1 \varphi _0 \left( {\begin{array}{*{20}c}
   q^{-n}  \\
    -   \\
\end{array};q,z} \right)
 = 
	\frac{ (q^{-n} z;q)_\infty}{ (z;q)_\infty} = (q^{-n}z;q)_n,
\end{equation*}
and there are no longer any convergence conditions.

It will also prove convenient to have a notation for the $q$-binomial
coefficient:
\begin{equation*}
\qbinom{n}{k}_q  = \frac{ (q;q)_n}{(q;q)_k (q;q)_{n-k}}
\end{equation*}
and for the $q$-numbers:
\begin{equation*}
[a]_q = \frac{ 1-q^a}{1-q},\quad q\neq 1.
\end{equation*}

\section{The Wigner distribution function}

The Wigner function for stationary states of a quantum system can be obtained 
from its wave functions $\psi_n(x)$ or $\tilde \psi_n(p)$ 
(in the position or momentum representation) by the well-known definition~\cite{hillery}
\numparts
\begin{eqnarray}
\label{2a}
W_n (p,x) = \frac{1}{{2\pi \hbar}}\int\limits_{ - \infty }^\infty  {\psi {}_n^* \left( {x - {\textstyle{1 \over 2}}x'} \right)\psi {}_n\left( {x + {\textstyle{1 \over 2}}x'} \right)e^{ - ipx'/\hbar} \rmd x'}, 
\\
\label{2b}
W_n (p,x) = \frac{1}{{2\pi \hbar}}\int\limits_{ - \infty }^\infty  {\tilde \psi {}_n^* \left( {p - {\textstyle{1 \over 2}}p'} \right)\tilde \psi {}_n\left( {p + {\textstyle{1 \over 2}}p'} \right)e^{ixp'/\hbar} \rmd p'}. 
\end{eqnarray}
\endnumparts

It allows one to calculate the quantum average of a physical quantity $f$ by the formula
\begin{equation}
\label{3}
\bar f_n  = \int\limits_{ - \infty }^\infty \int\limits_{ - \infty }^\infty  {f(p,x)W_n (p,x)\rmd p \rmd x}, 
\end{equation}
where $f(p,x)$ is the Weyl symbol of the operator $\hat f(\hat p,\hat x)$~\cite{tatarskii}.

The following explicit expression of the Wigner function for the stationary states of the non-relativistic linear harmonic oscillator is well-known~\cite{tatarskii,davies}: 
\begin{equation}
\label{4}
\fl \quad W_n ^{HO} \left( {p,x} \right) = \frac{{\left( { - 1} \right)^n }}{{\pi \hbar }}\exp \left[ { - \frac{2}{{\hbar \omega }}\left( {\frac{{p^2 }}{{2m}} + \frac{{m\omega ^2 x^2 }}{2}} \right)} \right] \cdot L_n \left( {\frac{4}{{\hbar \omega }}\left( {\frac{{p^2 }}{{2m}} + \frac{{m\omega ^2 x^2 }}{2}} \right)} \right),
\end{equation}
where $L_n$ are the Laguerre polynomials~\cite[Section 1.11]{KoeSwart}.

The Wigner distribution function determined in such a way takes both negative and positive values, therefore it is only a quasiprobability distribution function of $p$ and $x$. It is bounded by the restriction
$\left| {W_n \left( {p,x} \right)} \right| \le \left( {\pi \hbar } \right)^{ - 1}$~\cite{royer}. For this reason a class of Gaussian smoothed distribution functions with nonnegative behaviour was introduced~\cite{cartwright,janssen,rajagopal1}:
\begin{equation}
\label{5}
\fl \quad \overline W_n \left( {p,x} \right) = \frac{1}{{\pi \hbar }}\int\limits_{ - \infty }^\infty  {\int\limits_{ - \infty }^\infty  {\exp \left( { - \frac{{{p'} ^2 }}{{2\Delta _p ^2 }} - \frac{{{x'} ^2 }}{{2\Delta _x ^2 }}} \right) \cdot W_n \left( {p + p' ,x + x' } \right)\rmd p' \rmd x' } }  \ge 0,
\end{equation}
where $\Delta _p \Delta _x $ is a finite region of the phase plane. The simplest form of the Gauss smoothed Wigner function is when $\Delta _p \Delta _x =\hbar/2$, where (\ref{5}) will be a so-called 
Husimi distribution function~\cite{husimi}:
\begin{equation}
\label{6}
\fl \quad \overline W_n \left( {p,x} \right) = \frac{1}{{\left( {2\pi } \right)^{3/2} \hbar \Delta _x }}\left| \; {\int\limits_{ - \infty }^\infty  {\psi _n \left( {x'} \right) \cdot \exp \left[ { - \frac{{ipx'}}{\hbar } - \frac{{\left( {x - x'} \right)^2 }}{{4\Delta _x ^2 }}} \right]\rmd x'} } \right|^2 .
\end{equation}

Equation~(\ref{6}) has the property that $\overline W_n \left( {p,x} \right)$ is restricted by the condition $0 \le \overline W_n \left( {p,x} \right) \le \left( {\pi \hbar } \right)^{ - 1} $. 

The calculation of the distribution function of the non-relativistic harmonic oscillator with this formula, when the parameter $\Delta _x ^2$ is taken equal to $\hbar/2m \omega$ (simplest case), leads to the expression~\cite{tatarskii}:
\begin{equation}
\label{7}
\fl \quad \overline W_n ^{HO} \left( {p,x} \right) = \left( {2\pi \hbar n!} \right)^{ - 1} \left[ {\frac{1}{{\hbar \omega }}\left( {\frac{{p^2 }}{{2m}} + \frac{{m\omega ^2 x^2 }}{2}} \right)} \right]^n  \cdot \exp \left[ { - \frac{1}{{\hbar \omega }}\left( {\frac{{p^2 }}{{2m}} + \frac{{m\omega ^2 x^2 }}{2}} \right)} \right].
\end{equation}

\section{The $q$-deformed harmonic oscillator model}

The $q$-deformed harmonic oscillator model considered in this
article was already developed in a number of papers~\cite{macfarlane,kagramanov,atakishiyev,mirkasimov,vanderjeugt,atakishiyev2,atakishiyev3,nagiyev,atakishiyev4}. 
It has the following creation and annihilation operators:
\begin{equation}
\label{eq:b+-}
b^{\pm} = \pm \frac{i}{\sqrt{1-q}} \mexp{\mp\lambda x^2}\bigl(
\mexp{\mp 2i\lambda h x} - q^{1/2}\mexp{-\frac{ih}{2}\partial_x}\bigr)
\mexp{\pm \lambda x^2},
\end{equation}
with $\lambda = \frac{m\omega}{2\hbar}$. Herein, $h$ is a deformation parameter related to a finite-difference method with respect to $x$, and
\begin{equation*}
q=\mexp{-\lambda h^2}.
\end{equation*}

In the rest of this article, $q$ will always have the value $\mexp{-\lambda h^2}$. 

Note that an operator of the form 
$\exp(a\partial_x)$, where $a$ is an arbitrary complex number,
acts as follows on functions of $x$:
\begin{equation*}
\exp(a\partial_x) f(x) = f(x+a).
\end{equation*}

In the limit $h \to 0$ (or $q \to 1$) the operators (\ref{eq:b+-}) reduce to the well-known non-relativistic harmonic oscillator creation and annihilation operators~\cite{nagiyev}:
\begin{equation*}
\mathop {\lim }\limits_{h \to 0} b^ \pm  = a^ \pm   =  \mp \frac{1}{{2\sqrt \lambda  }}e^{ \pm \lambda x^2 } \frac{\partial }{{\partial x}}e^{ \mp \lambda x^2 } .
\end{equation*}

It is easily verified that
\begin{equation*}
[b^-,b^+]_q = b^-b^+ - q b^+b^- = 1.
\end{equation*}

The Hamiltonian operator $H$ of the model is then
\begin{equation*}
H = \hbar \omega \left( {\sqrt q \; b^ +  b^ -   + \left[ {{\textstyle{1 \over 2}}} \right]_q} \right).
\end{equation*}

The wave functions for this model in the $x$-representation~\cite{atakishiyev5} are related
to the Rogers-Szeg\"o polynomials 
$\mathcal{H}_n(x\,\vert\, q)$~\cite[Chapter 17]{Ismail}.  More in 
particular, one has
\begin{equation}
\label{eq:wave-x}
\psi_n^{qHO}\left(x \right) = c_n 
\mathcal{H}_n\left( -{\mexp{-2i\lambda hx}} \,\vert\, q\right)\mexp{-\lambda x^2 },
\end{equation}
where the normalization constant $c_n$ is given by:
\begin{equation*}
c_n = \left({\frac{2\lambda}{\pi}}\right)^{1/4} (-i)^n q^{n/2} (q;q)_n^{-1/2}.
\end{equation*}

One can see that in the limit for $h {\to} 0$ one recovers the wave functions for the non-relativistic quantum mechanical harmonic oscillator in the $x$-representation:
\begin{equation}
\label{wave-ho}
\psi _n ^{qHO} \left( x \right)\mathop  \to \limits^{h \to 0} \psi _n ^{HO} \left( x \right) = \frac{1}{{\sqrt {2^n n!\sqrt {\pi /2\lambda } } }} H_n \left( {\sqrt {2\lambda } \; x } \right) \cdot e^{ - \lambda x^2 }.
\end{equation}

The above limit can be verified by employing the following known limit relation between the Rogers-Szeg\"o and Hermite polynomials~\cite{nagiyev,atakishiyev5}:
\begin{equation}
\label{rs->her}
\mathop {\lim }\limits_{\alpha \stackrel{>}\to 0} \left( { - i\sqrt {\frac{{2\tilde q}}{{1 - \tilde q}}} } \right)^n \mathcal{H}_n \left( { - e^{ - 2i\alpha y} |\tilde q} \right) = H_n \left( y \right),\quad \tilde q = e^{ - 2\alpha ^2 } ,\quad \alpha  > 0.
\end{equation}

The wave functions in the momentum and space representation are related through a Fourier transform:
\begin{equation*}
\tilde\psi_n^{qHO}(p) = \frac{1}{\sqrt{2\pi\hbar}} \int_{-\infty}^\infty
\psi_n^{qHO}(x)\mexp{-\frac{ixp}{\hbar}}\, \rmd x. 
\end{equation*}

Those in the $p$-representation are expressed through the Stieltjes-Wigert polynomials $S_n\left(x;q \right)$:
\begin{equation}
\label{eq:wave-p}
	\tilde\psi_n^{qHO}(p) = \frac{c_n (q;q)_n}{\sqrt{2\lambda\hbar}} 
	S_n\left(q^{-1/2} \mexp{-\frac{hp}{\hbar}};q\right)
	\mexp{-\frac{p^2}{4\lambda\hbar^2}},
\end{equation}
where we have used the notation and normalization 
of~\cite[Section 3.27]{KoeSwart} for 
the Stieltjes-Wigert polynomials $S_n\left(x;q \right)$.

Also here, one recovers the correct expression of the wave functions for the non-relativistic quantum mechanical harmonic oscillator in the $p$-representation when $h\rightarrow 0$:
\begin{equation}
\label{wave-ho-p}
\tilde\psi_n^{qHO} \left( p \right)\mathop  \to \limits^{h \to 0} \tilde \psi _n ^{HO} \left( p \right) = \frac{\left(-i \right)^n}{{\sqrt {2^n n!\sqrt {\pi \cdot m \omega \hbar } } }} H_n \left( {p/\sqrt {m \omega \hbar } } \right) \cdot e^{ - \frac{p^2}{2m \omega \hbar} },
\end{equation}
which follows from the limit relation between the Stieltjes-Wigert and Hermite polynomials~\cite{nagiyev,atakishiyev5}:
\begin{equation}
\label{sw->her}
\fl \quad \mathop {\lim }\limits_{\stackrel{>} {\alpha  \to 0}} \left(  \frac{2\tilde q}{1 - \tilde q} \right)^{n/2} \left( {\tilde q;\tilde q} \right)_n S_n \left( { {\tilde q}^{-1/2} e^{ - 2\alpha y} ;\tilde q} \right) = H_n \left( y \right),\quad \tilde q = e^{ - 2\alpha ^2 } ,\quad \alpha  > 0.
\end{equation}

Thanks to the orthogonality relations for the Rogers-Szeg\"o and Stieltjes-Wigert polynomials on the real axis~\cite{atakishiyev5},
\begin{eqnarray}
\fl \quad \frac{1}{{\sqrt \pi  }}\int\limits_{ - \infty }^\infty  {\mathcal{H}_n \left( { - e^{ - 2i\alpha y} | \tilde q} \right) \cdot \mathcal{H}_m \left( { - e^{ 2i\alpha y} | \tilde q} \right) \cdot e^{ - y^2 } \rmd y}  = \frac{{\left( {\tilde q; \tilde q} \right)_n }}{{{\tilde q}^n }}\delta _{nm}, \\
\fl \quad \frac{1}{{\sqrt \pi  }}\int\limits_{ - \infty }^\infty  {S_n \left( { {\tilde q}^{ - 1/2} e^{ - 2\alpha y} ;\tilde q} \right) \cdot S_m \left( { {\tilde q}^{ - 1/2} e^{ - 2\alpha y} ; \tilde q} \right) \cdot e^{ - y^2 } \rmd y}  = \frac{1}{{\left( {\tilde q; \tilde q} \right)_n } {{\tilde q}^n }}\delta _{nm} ,
\end{eqnarray}
both wave functions~\eref{eq:wave-x} and~\eref{eq:wave-p} are also orthonormal:
\begin{equation}
\fl \quad \int \limits_{-\infty}^\infty {\psi_n^*}^{qHO}(x)\psi_m^{qHO}(x)\, \rmd x = \delta_{nm}, \qquad
\int \limits_{-\infty}^\infty {{{\tilde \psi}^*}_n}\,^{qHO}(p)\tilde \psi_m^{qHO}(p)\, \rmd p = \delta_{nm}.
\end{equation}

Finally, let us mention that the spectrum of the Hamiltonian $H$ is given by:
\begin{equation*}
	E_{n,q} = \hbar\omega\left[n+\frac12 \right]_q,\quad \mbox{with}
	\quad n=0,1,2,\ldots
\end{equation*}

\section{Computation of distribution functions}
\subsection{The Wigner distribution function for stationary states}

The Wigner function for stationary states of the $q$-oscillator can be computed 
using~\eref{2a} and~\eref{eq:wave-x} or~\eref{2b} and~\eref{eq:wave-p}.  
Substitution of~\eref{eq:wave-x} in~\eref{2a} leads to the following integral:
\begin{equation}
\label{wig-int}
\fl \quad W_{n,q} \left( {p,x} \right) = \frac{{\left| {c_n } \right|^2 }}{{2\pi \hbar }} \int\limits_{ - \infty }^\infty  {\mathcal{H}_n \left( {\left. { - e^{i\lambda h\left( {2x - x'} \right)} } \right|q} \right)\mathcal{H}_n \left( {\left. { - e^{ - i\lambda h\left( {2x + x'} \right)} } \right|q} \right)\mexp{-\lambda(x-\frac{x'}{2})^2 - \lambda(x+\frac{x'}{2})^2-i p x'/\hbar} \rmd x'} .\;
\end{equation}

Here, the Rogers-Szeg\"o polynomials have the following explicit expression:
\begin{equation}
	\label{eq:Hermite-Szego}
	\mathcal{H}_n(x\,\vert\, \tilde q) = \sum_{k=0}^n 
	\qbinom{n}{k}_{\tilde q} 
		\left(\frac{x}{{\tilde q}^{1/2}}\right)^k,
		\quad\mbox{with}\quad
	0 < \tilde q < 1.
\end{equation}
Using this expression, one writes the Wigner function \eref{wig-int} 
explicitly as
\begin{equation*}
\eqalign{& W_{n,q}(p,x) = \frac{1}{2\pi\hbar} |c_n|^2 
\sum_{k,s=0}^n (-1)^{k+s} q^{-(k+s)/2} \qbinom{n}{k}_q
\qbinom{n}{s}_q  \\
&\quad\times 
\int_{-\infty}^\infty  
\mexp{-\lambda(x-\frac{x'}{2})^2 - \lambda(x+\frac{x'}{2})^2}
\mexp{2i\lambda h(x-\frac{x'}{2})k}
\mexp{-2i\lambda h(x+\frac{x'}{2})s}
\mexp{-\mfrac{ipx'}{\hbar}}\,\rmd x'.}
\end{equation*}
Next, we use the Gaussian integral:
\begin{equation}\label{eq:gauss}
\fl \quad	\int_{-\infty}^\infty \mexp{-(a_2\,x^2+a_1\, x + a_0)}\, \rmd x 
	= \sqrt{\frac{\pi}{a_2}}
	\mexp{-a_0+\mfrac{a_1^2}{4a_2}},\quad\mbox{when}\quad a_2 >0\ 
	\mbox{and}\  a_1,a_0\in\C.
\end{equation}
After some trivial simplifications, one then finds the following double
sum expression for the Wigner distribution function:
\begin{equation}
\label{wq-dsum}
\fl W_{n,q}(p,x)  = \frac{1}{\pi\hbar}\frac{q^n}{(q;q)_n}
	\mexp{-2\lambda x^2 - \mfrac{p^2}{2\lambda\hbar^2}} 
	\sum_{k,s=0}^n (-1)^{k+s}  
	\qbinom{n}{k}_q \qbinom{n}{s}_q
 q^{\binom{k}{2}+  \binom{s}{2}+ ks}\mexp{-ka^*-sa},
\end{equation}
with 
\begin{equation}
\label{a}
a=\frac{hp}{\hbar} + 2i\lambda h x.
\end{equation}

This expression can be further reduced to 
a single sum by applying the $q$-binomial theorem once.  Indeed, we have:
\begin{equation*}
\fl \eqalign{W_{n,q}(p,x)
&= \frac{1}{\pi\hbar}\frac{q^n}{(q;q)_n} 
\mexp{{ - \frac{2}{{\hbar \omega }}\left( {\frac{{p^2 }}{{2m}} + \frac{{m\omega ^2 x^2 }}{2}} \right)}}
	\sum_{s=0}^n (-1)^{s}  
	\qbinom{n}{s}_q 
 q^{\binom{s}{2} }\mexp{-sa} 
	\, _1 \varphi _0 \left( {\begin{array}{*{20}c}
   {q^{ - n} }  \\
    -   \\
\end{array};q,q^{s + n} e^{ - a^* } } \right)\\
	&= \frac{1}{\pi\hbar}\frac{q^n}{(q;q)_n} 
	\mexp{{ - \frac{2}{{\hbar \omega }}\left( {\frac{{p^2 }}{{2m}} + \frac{{m\omega ^2 x^2 }}{2}} \right)}}
	\sum_{s=0}^n (-1)^{s}  
	\qbinom{n}{s}_q 
 q^{\binom{s}{2}  } \mexp{-sa}\, (q^s\mexp{-a^*};q)_n\\
	&= \frac{1}{\pi\hbar}\frac{q^n}{(q;q)_n} 
	\mexp{{ - \frac{2}{{\hbar \omega }}\left( {\frac{{p^2 }}{{2m}} + \frac{{m\omega ^2 x^2 }}{2}} \right)}}
	(\mexp{-a^*};q)_n\, _2 \varphi _1 \left( {\begin{array}{*{20}c}
   {q^{ - n} ,q^n e^{ - a^* } }  \\
   {e^{ - a^* } }  \\
\end{array};q,q^n e^{ - a} } \right) .}
\end{equation*}	

Applying now the following transformation formula~\cite[III.7]{GR2}:
\begin{equation*}
_2 \varphi _1 \left( {\begin{array}{*{20}c}
   {\tilde q^{ - n} ,b}  \\
   c  \\
\end{array};\tilde q,z} \right) = 
\frac{(c/b; \tilde q)_n}{(c; \tilde q)_n}\,   
_3 \varphi _2 \left( {\begin{array}{*{20}c}
   {\tilde q^{ - n} ,b,bz\tilde q^{ - n} /c}  \\
   {b\tilde q^{1 - n} /c,0}  \\
\end{array};\tilde q,\tilde q} \right),
\end{equation*}
leads to the following expression
\begin{equation}
	\fl \quad W_{n,q}(p,x)  =\frac{(-1)^n}{\pi\hbar}q^{-\binom{n}{2}} 
\mexp{{ - \frac{2}{{\hbar \omega }}\left( {\frac{{p^2 }}{{2m}} + \frac{{m\omega ^2 x^2 }}{2}} \right)}}\, 
_3 \varphi _2 \left( {\begin{array}{*{20}c}
   {q^{ - n} ,q^n e^{ - a} ,q^n e^{ - a^* } }  \\
   {q,0}  \\
\end{array};q,q} \right).
\label{3phi2}
\end{equation}

Using the classical definition of the Al-Salam-Chihara polynomials~\cite[Section 3.8]{KoeSwart}
\begin{equation*}
Q_n \left( {y;\alpha,\beta \left| {\tilde q} \right.} \right) = \frac{{\left( {\alpha \beta ;\tilde q} \right)_n }}{{\alpha ^n }}{}_3\varphi _2 \left( {\begin{array}{*{20}c}
   {{\tilde q}^{ - n} , \alpha \; e^{i\theta } ,\alpha \; e^{-i\theta } }  \\
   {\alpha \beta,0}  \\
\end{array};\tilde q,\tilde q} \right),\;y = \cos \theta ,
\end{equation*}
the Wigner function~\eref{3phi2} can formally be written as
\begin{equation}
\label{wq-als}
\fl \quad W_{n,q} \left( {p,x} \right) = \frac{{\left( { - 1} \right)^n }}{\pi \hbar} \frac{q^{n\left(n+1 \right)/2}}{{\left( {q;q} \right)_n }} \cdot e^{ - n\frac{h}{\hbar }p}  \cdot e^{ - \frac{2}{{\hbar \omega }}\left( {\frac{{p^2 }}{{2m}} + \frac{{m\omega ^2 x^2 }}{2}} \right)}  \cdot Q_n \left( {\cos 2\lambda hx;q^n e^{ - \frac{h}{\hbar }p} ,q^{1 - n} e^{\frac{h}{\hbar }p} \left| q \right.} \right).
\end{equation}

Quite generally, Wigner functions for stationary states satisfy the following relation~\cite{hillery}:
\begin{equation}
\label{wigq-ort}
\int\limits_{ - \infty }^\infty  {\int\limits_{ - \infty }^\infty  {W_{n,q} \left( {p,x} \right) \cdot W_{m,q} \left( {p,x} \right)\rmd x} \rmd p}  = \frac{{\delta _{nm} }}{{2\pi \hbar }}.
\end{equation}

To prove this relation in the current case, one can use the double sum expression of the 
Wigner function~\eref{wq-dsum}. After simple calculations, using the Gaussian integral~\eref{eq:gauss}, the computation reduces to checking that
\begin{equation}
\label{wigq-ort-sum}
\fl \quad \frac{{q^{n + m} }}{{\left( {q;q} \right)_n \left( {q;q} \right)_m }}\sum\limits_{k = 0}^n {\left( {q^n } \right)^k \frac{{\left( {q^{ - n} ;q} \right)_k }}{{\left( {q;q} \right)_k }}} \left( {q^{ - k} ;q} \right)_m  \cdot \sum\limits_{k' = 0}^m {\left( {q^m } \right)^{k'} \frac{{\left( {q^{ - m} ;q} \right)_{k'} }}{{\left( {q;q} \right)_{k'} }}} ( {q^{ - k'} ;q} )_n  = \delta _{nm} .
\end{equation}

This is now readily verified since, in the left hand side, all terms except the one with $k=k'=m=n$ are zero.

From the ${}_3\varphi_2$ expression~\eref{3phi2} for the Wigner distribution function,
it is easy to verify that in the limit for $h\stackrel{>}{\to}0$ one 
recovers the Wigner distribution function for the 
non-relativistic quantum harmonic oscillator. 

First, using series expansion in $h$ one sees easily that 
\begin{equation*}
\eqalign{
& \lim_{\stackrel{>} {h\to 0}} \frac{(q^{n}\mexp{-a};q)_k(q^{n}\mexp{-a^*};q)_k}{(q;q)_k} 
 = \prod_{j=0}^{k-1}\lim_{\stackrel{>} {h\to 0}}
\frac{ (1-q^{n+j}\mexp{-a})(1-q^{n+j}\mexp{-a^*})}{1-q^{j+1}} \\
& \quad = \prod_{j=0}^{k-1}\lim_{\stackrel{>} {h\to 0}}
\left(\frac{{ \frac{4}{{\hbar \omega }}\left( {\frac{{p^2 }}{{2m}} + \frac{{m\omega ^2 x^2 }}{2}} \right)}}{j+1} + \mathcal{O}(h) \right)
 = \frac{\left(\frac{4}{{\hbar \omega }}\right)^k \left( {\frac{{p^2 }}{{2m}} + \frac{{m\omega ^2 x^2 }}{2}} \right)^k}{k!}.}
\end{equation*}

Secondly, using a term wise limit for
the ${}_3\varphi_2$-series, one finds that:
\begin{equation*}
\fl \quad \eqalign{
\lim_{\stackrel{>}{h\to 0}} W_{n,q}(p,x) & = 
\frac{(-1)^n}{\pi\hbar}\mexp{{ - \frac{2}{{\hbar \omega }}\left( {\frac{{p^2 }}{{2m}} + \frac{{m\omega ^2 x^2 }}{2}} \right)}}
\sum_{k=0}^n 
\lim_{\stackrel{>}{h\to 0}} \frac{(q^{-n};q)_k}{(q;q)_k} 
\frac{(q^{n}\mexp{-a};q)_k(q^{n}\mexp{-a^*};q)_k}{(q;q)_k} q^k
\\
& = \frac{(-1)^n}{\pi\hbar}\mexp{{ - \frac{2}{{\hbar \omega }}\left( {\frac{{p^2 }}{{2m}} + \frac{{m\omega ^2 x^2 }}{2}} \right)}}
\sum_{k=0}^n \frac{(-n)_k}{k!} \frac{\left(\frac{4}{{\hbar \omega }}\right)^k \left( {\frac{{p^2 }}{{2m}} + \frac{{m\omega ^2 x^2 }}{2}} \right)^k}{k!} \\
&=  \frac{(-1)^n}{\pi\hbar}\mexp{{ - \frac{2}{{\hbar \omega }}\left( {\frac{{p^2 }}{{2m}} + \frac{{m\omega ^2 x^2 }}{2}} \right)}}
\, L_n \left( {\frac{4}{{\hbar \omega }}\left( {\frac{{p^2 }}{{2m}} + \frac{{m\omega ^2 x^2 }}{2}} \right)} \right) = W_n^{HO}(p,x).}
\end{equation*}

\subsection{The Husimi distribution function}

The calculation of the Husimi distribution function~\eref{6} follows more
or less the same lines as that of the calculation of the Wigner distribution
function.

First, one computes:
\begin{equation}
\label{eq:Husimi}
\int_{-\infty}^\infty \psi_n(x')
\mexp{ -\mfrac{ipx'}{\hbar} - \mfrac{(x-x')^2}{4\Delta_x^2}} \, \rmd x'
\end{equation}
using the expressions~\eref{eq:wave-x},~\eref{eq:Hermite-Szego} and the Gaussian 
integral~\eref{eq:gauss}. Then, one can apply the $q$-binomial theorem.
Secondly, one conjugates this expression and multiplies both expressions, taking
into account the constant factor in~\eref{6}.  Finally, one  arrives 
at:
\begin{equation}
\label{hus-f}
	\overline W_{n,q}(p,x) = \frac{1}{2\pi\hbar} 
	q^n \frac{(\mexp{-a/2};q)_n (\mexp{-a^*/2};q)_n}{(q;q)_n}
	\mexp{{ - \frac{1}{{\hbar \omega }}\left( {\frac{{p^2 }}{{2m}} + \frac{{m\omega ^2 x^2 }}{2}} \right)}},
\end{equation}
with $a$ given by \eref{a}.

Using the same kind of limit calculation as before, one sees that the 
limit $h\stackrel{>}{\to}0$ (or $q\stackrel{<}{\to}1$) yields
\begin{equation*}
\lim_{h\stackrel{>}{\to}0} \overline W_{n,q}(p,x) = 
\frac{1}{2\pi\hbar\, n! } \left[{\frac{1}{{\hbar \omega }}\left( {\frac{{p^2 }}{{2m}} + \frac{{m\omega ^2 x^2 }}{2}} \right)}\right]^n 
\mexp{{ - \frac{1}{{\hbar \omega }}\left( {\frac{{p^2 }}{{2m}} + \frac{{m\omega ^2 x^2 }}{2}} \right)}}, 
\end{equation*}
which is indeed the correct expression for the Husimi distribution 
function \eref{7}.

\section{Discussions}

The explicit expressions obtained for the Wigner and Husimi distribution function for the $q$-deformed harmonic oscillator 
allow us to explore in detail the behaviour of these distribution functions in the phase space as $q$ varies. 
First, an analysis of the ground state ($n=0$, vacuum state) shows that the behaviour of the non-relativistic harmonic oscillator and its $q$-generalization in the phase space is the same. 
In other words, there is no impact of the parameter $q$ in the absence of photons, and the Wigner function of the ground state for the $q$-deformed oscillator is the well-known Gaussian centered at the phase space point $(0,0)$. 

\begin{figure}
\begin{center}
\begin{tabular}{cc}
\includegraphics[width=0.45\textwidth]{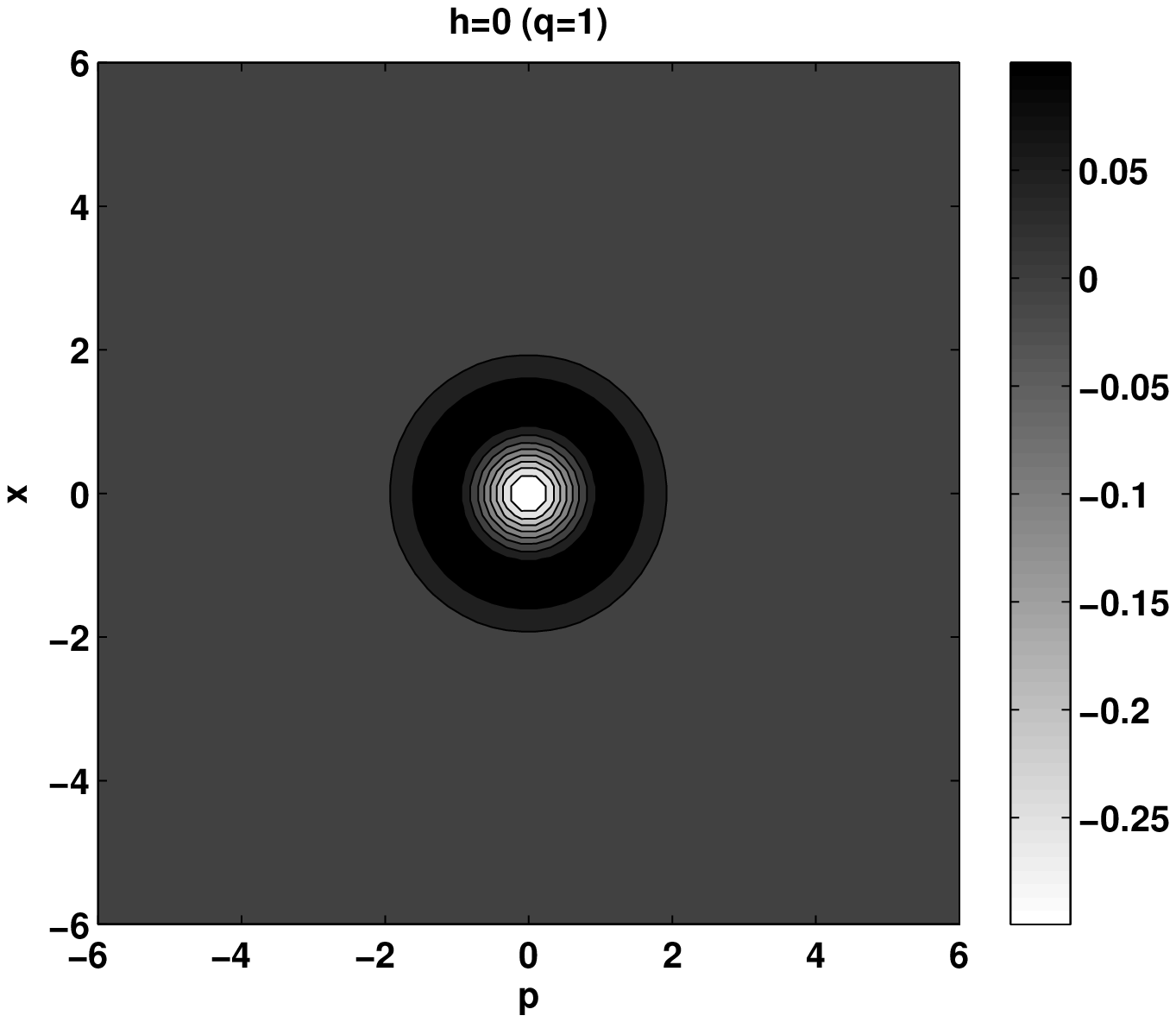}&
\includegraphics[width=0.45\textwidth]{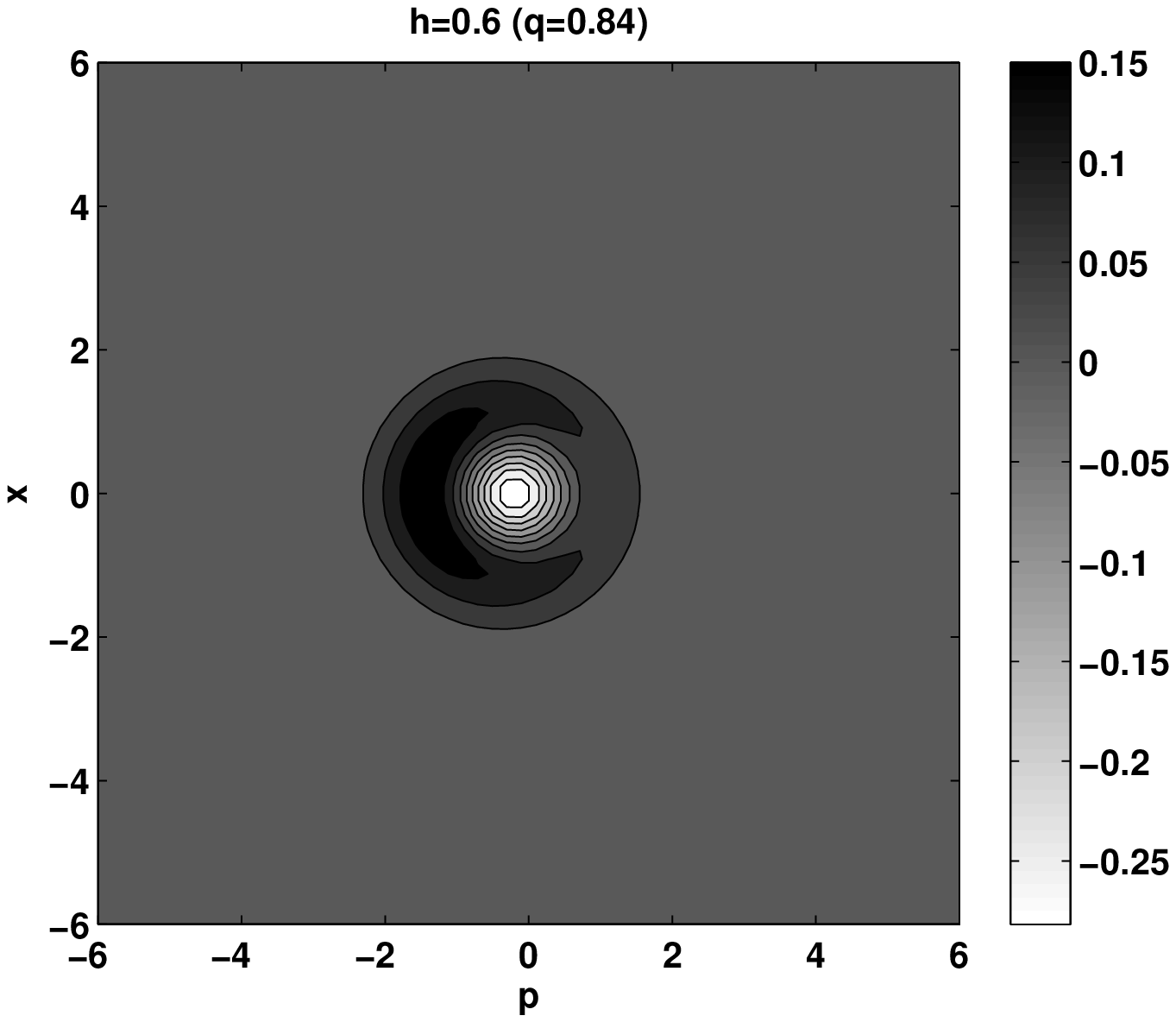}\\
\includegraphics[width=0.45\textwidth]{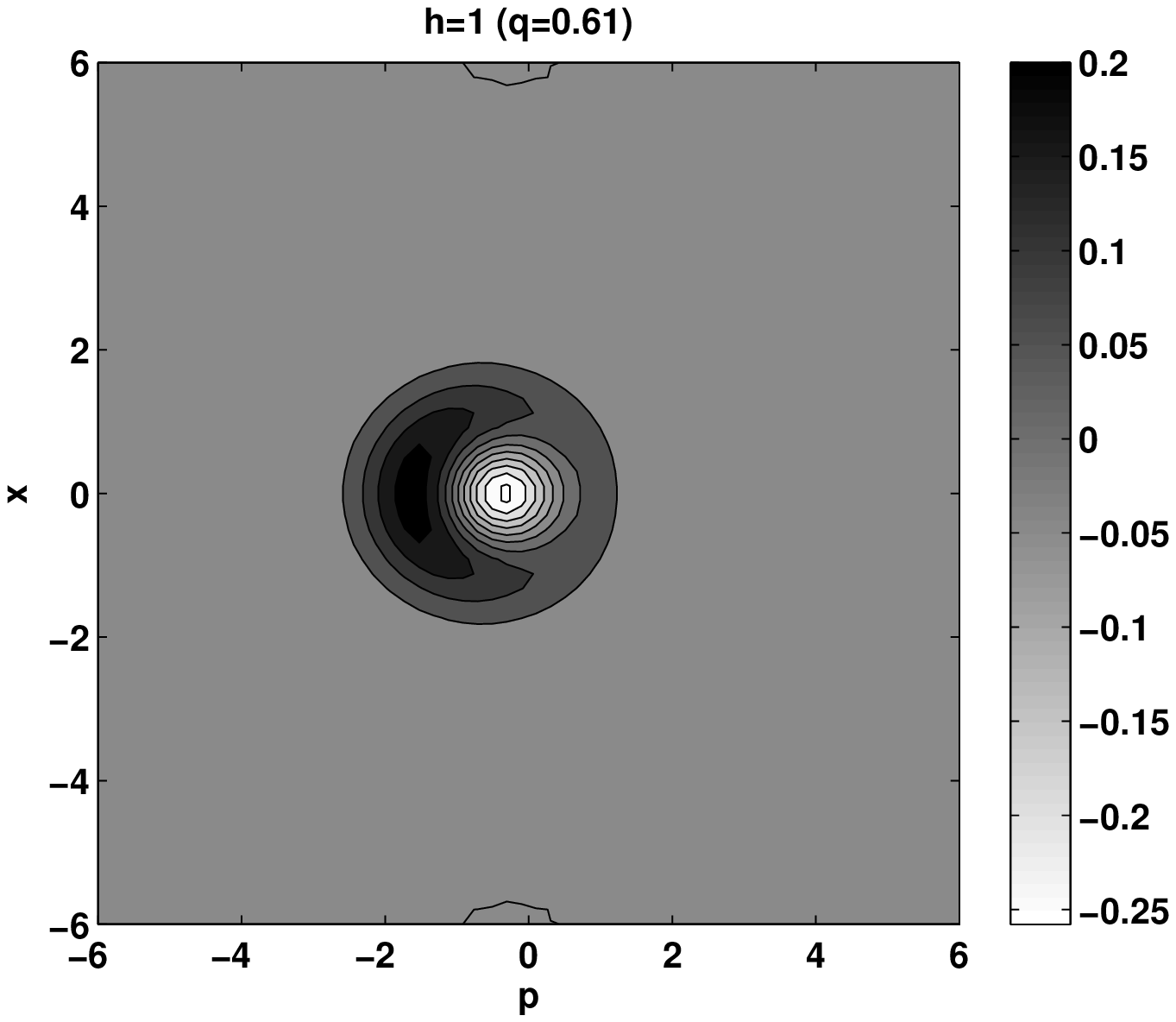}&
\includegraphics[width=0.45\textwidth]{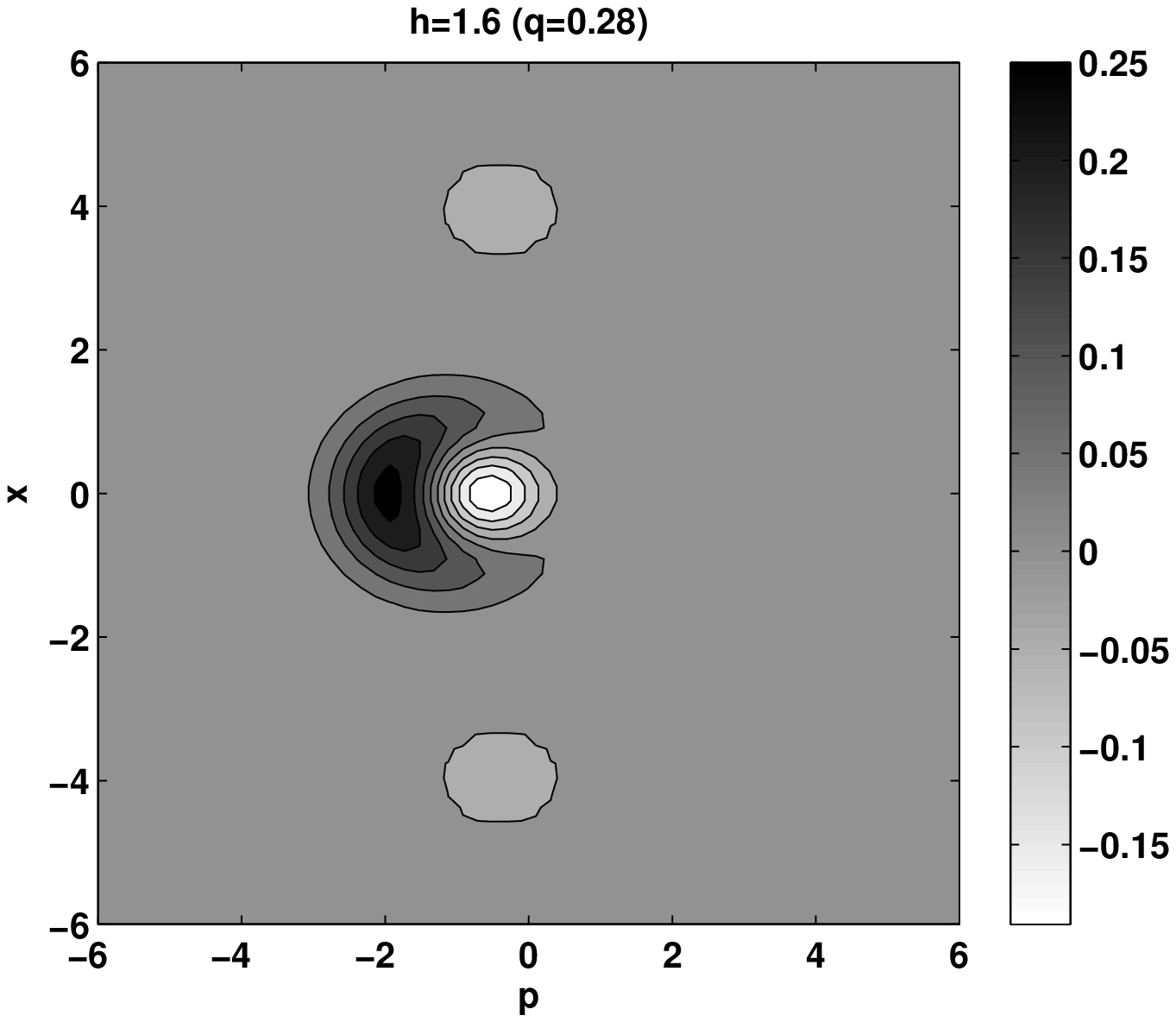}\\
\includegraphics[width=0.45\textwidth]{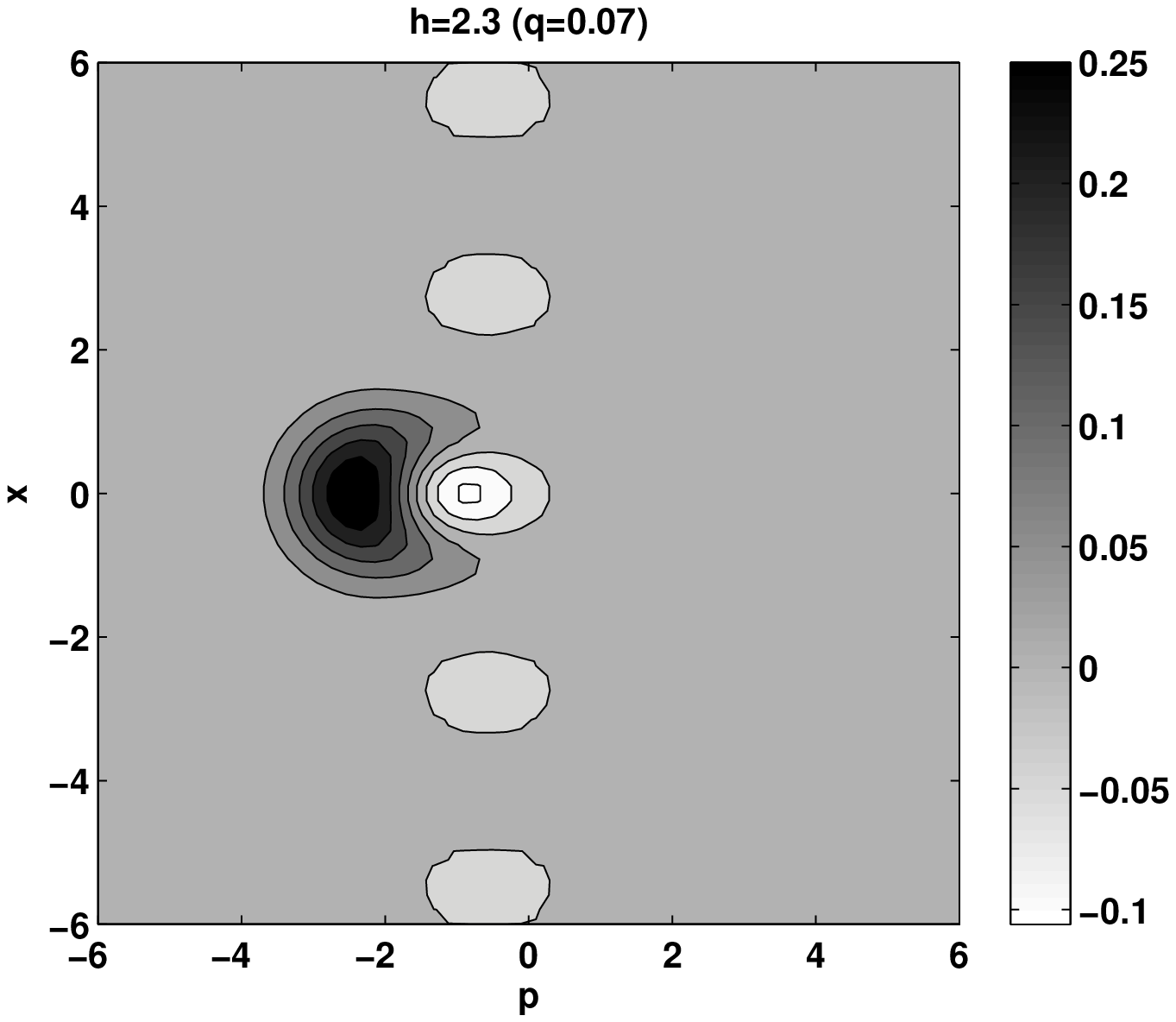}&
\includegraphics[width=0.45\textwidth]{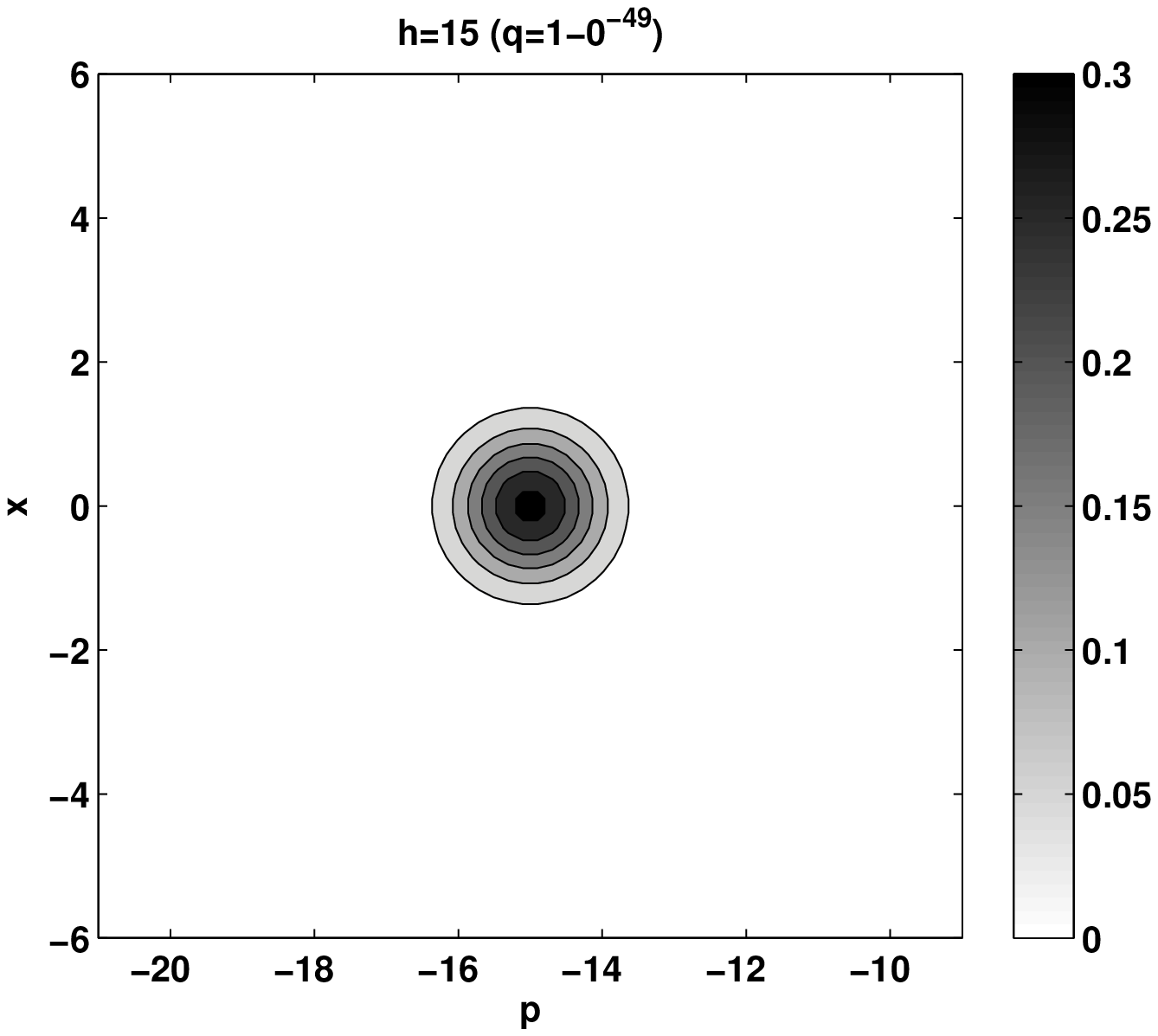}
\end{tabular}
\end{center}
\caption{A density plot of the Wigner function of the single photon state for the $q$-deformed harmonic oscillator, for values of $h=0;\;0.6;\;1;\;1.6;\;2.3;\;15$ ($q=1;\;0.84;\;0.61;\;0.28;\;0.07;\;10^{-49}$) and $m=\omega=\hbar=1$. The function changes dramatically when the value of the parameter $q$ decreases. Observe also a displacement of the Wigner function due to the non-zero average of the $q$-oscillator momentum.}
\end{figure}

\begin{figure}
\begin{center}
\begin{tabular}{cc}
\includegraphics[width=0.45\textwidth]{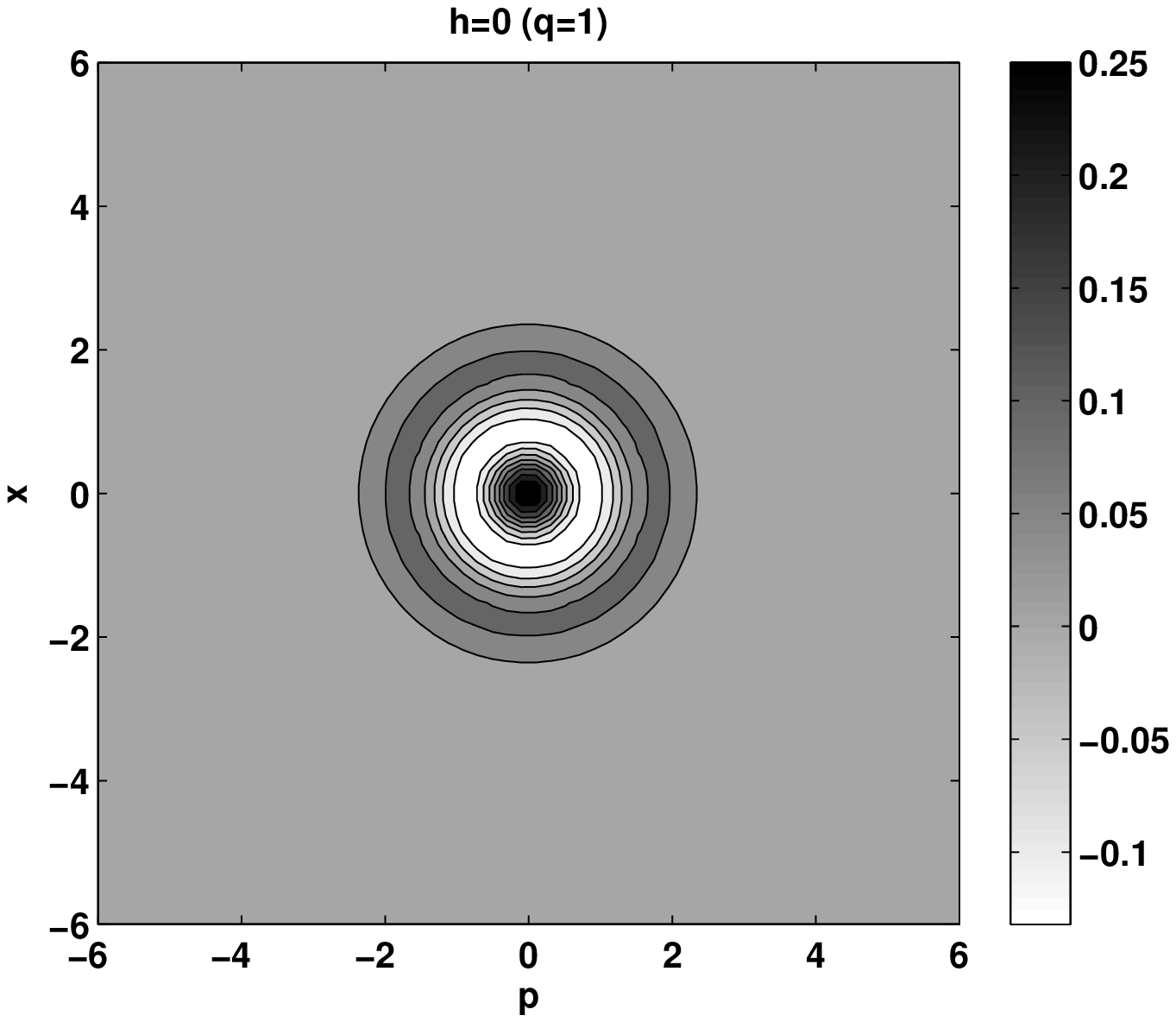}&
\includegraphics[width=0.45\textwidth]{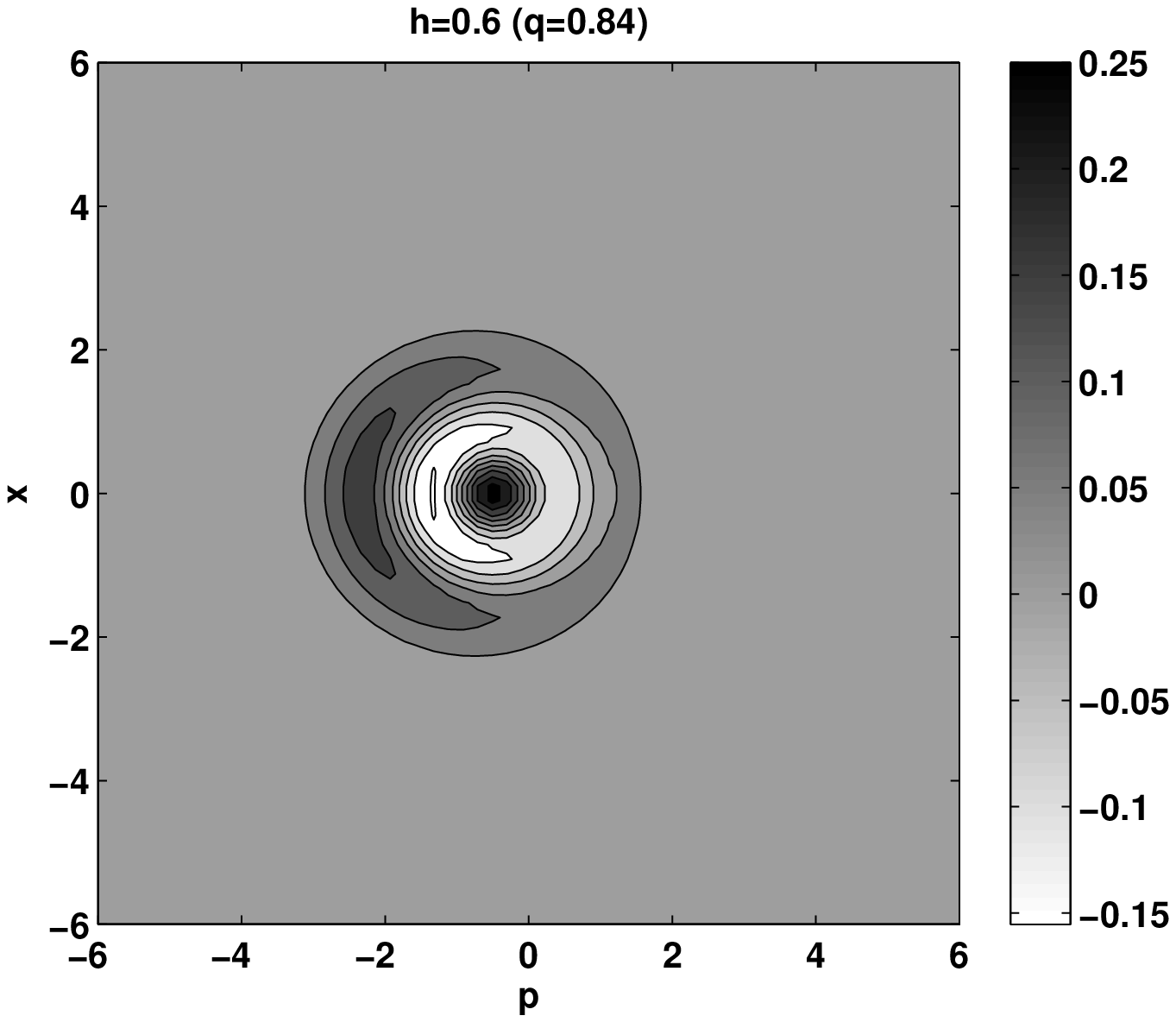}\\
\includegraphics[width=0.45\textwidth]{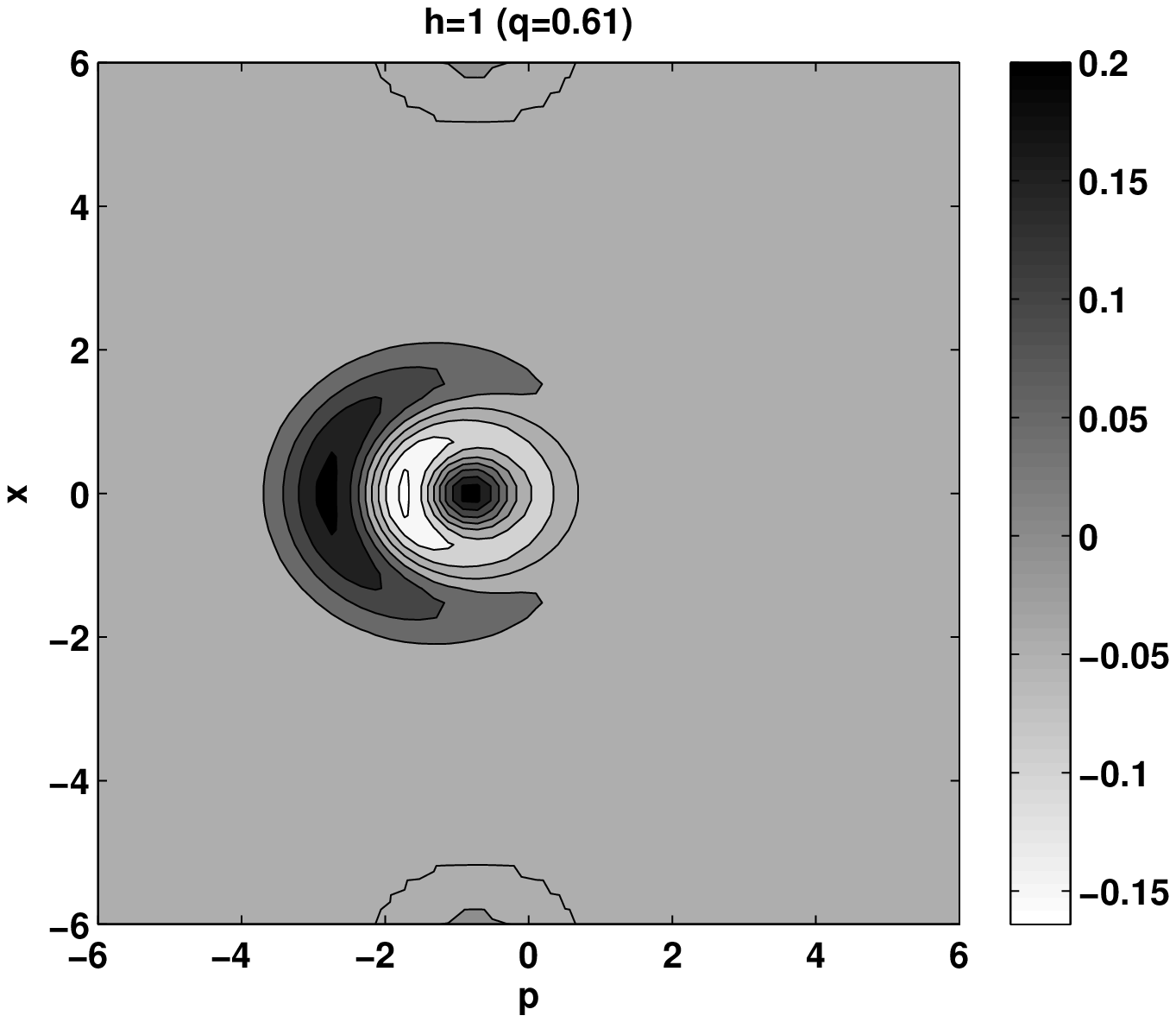}&
\includegraphics[width=0.45\textwidth]{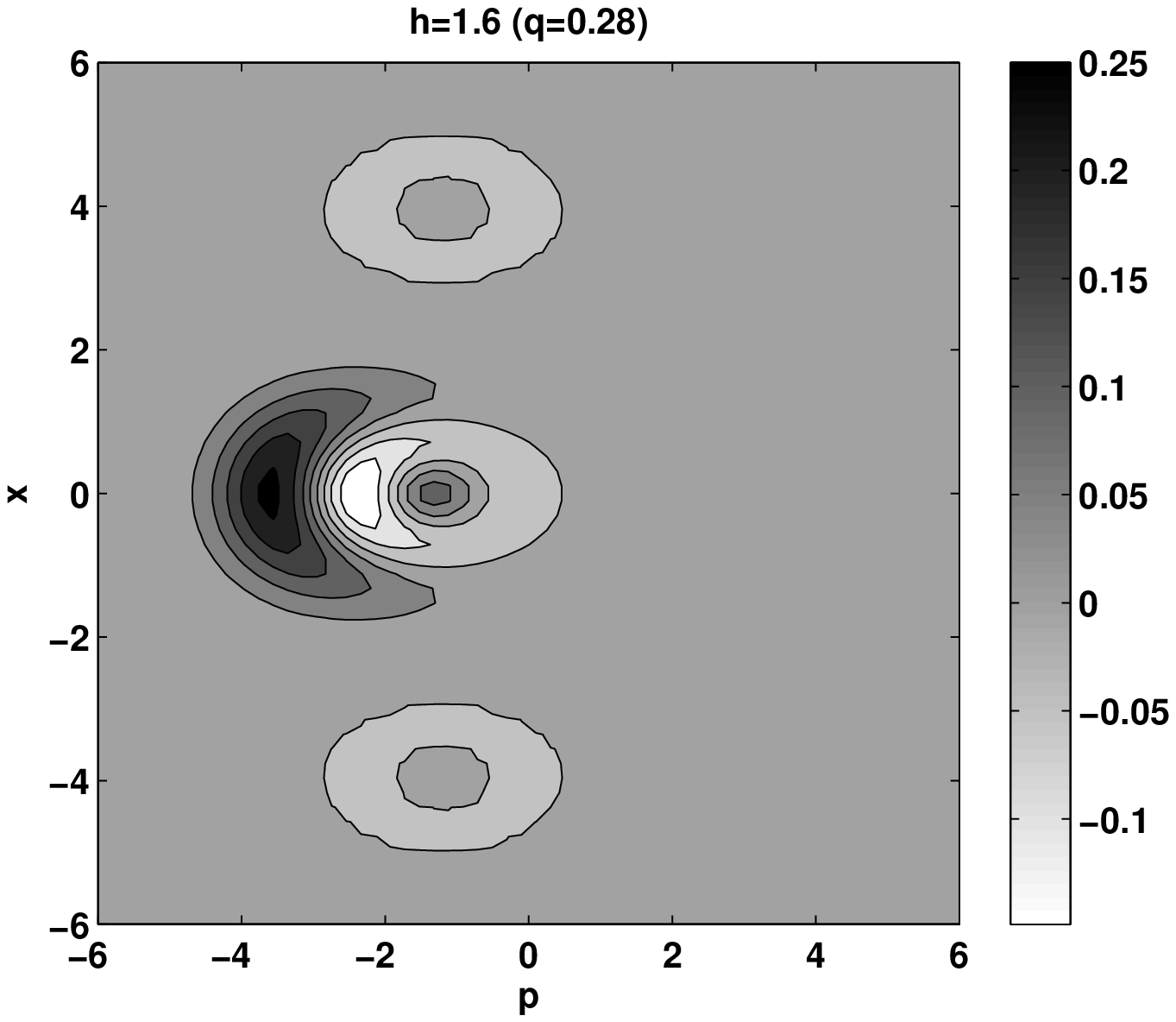}\\
\includegraphics[width=0.45\textwidth]{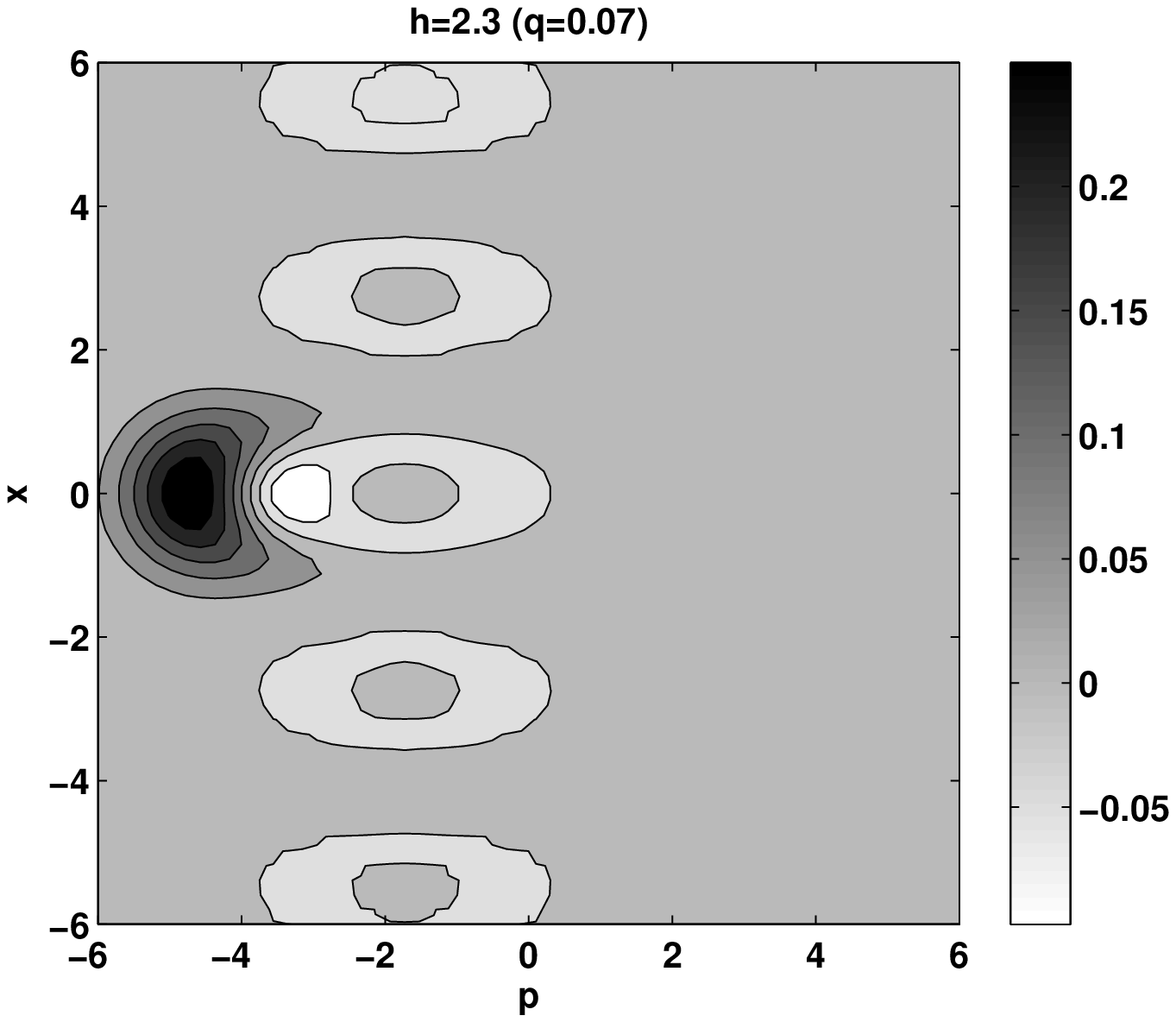}&
\includegraphics[width=0.45\textwidth]{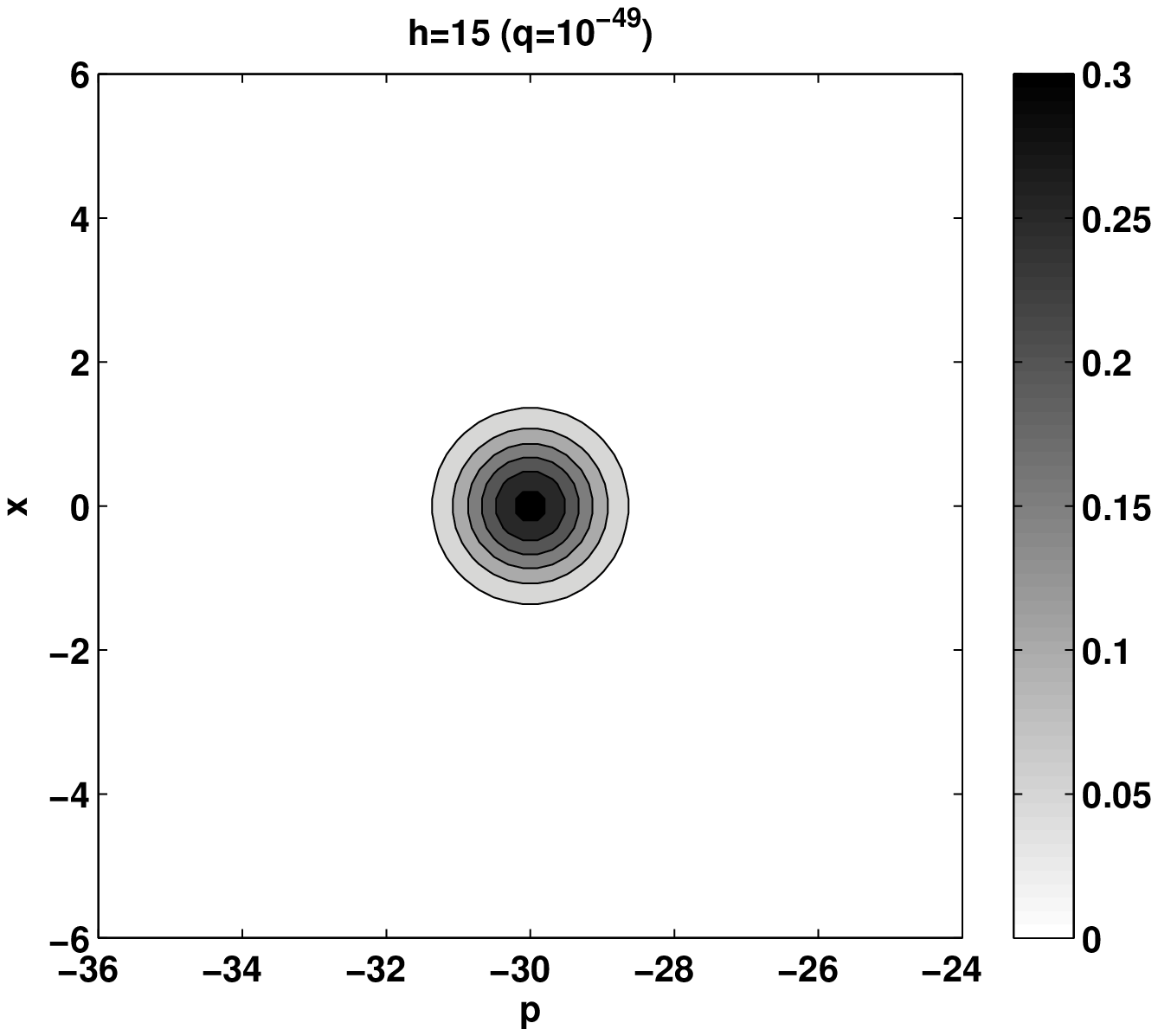}
\end{tabular}
\end{center}
\caption{A density plot of the Wigner function of the double photon state for the $q$-deformed harmonic oscillator, for values of $h=0;\;0.6;\;1;\;1.6;\;2.3;\;15$ ($q=1;\;0.84;\;0.61;\;0.28;\;0.07;\;10^{-49}$) and $m=\omega=\hbar=1$. The function changes again dramatically when the value of the parameter $q$ decreases. The displacement of the Wigner function due to the non-zero average of the $q$-oscillator momentum is clear, and one can also see the impact
of the momentum average dependence on~$n$.}
\end{figure}

For the excited quantum states the situation is rather different.
In Figs.~1 and~2 we have given a density plot for the single ($n=1$) and double ($n=2$) photon states of the $q$-deformed harmonic oscillator in the phase space. 
For simplicity, we use the scale $m=\omega=\hbar=1$. 
The first plot in each of the figures is the Wigner function for the non-relativistic quantum harmonic oscillator ($h=0$ or $q=1$). 
As long as $h\leq 1$ we find a phase space that is not so different from that of the non-relativistic harmonic oscillator, except for the occurrence of a new peak at the point $x=0$ and $p<0$. 
This value $h=1$ is important, because most of the investigations of finite-difference generalizations of the Schr\"odinger equation propose to take a finite-difference step $h$ to be equal to $\lambdabar=\hbar/mc$, i.e.\ the Compton wavelength of a particle with mass $m$ (see, for example~\cite{kagramanov,wall,march}). 
In this case ($h=1$) one has $\hbar \omega = m c^2$. 
After this point, as $h$ increases ($h>1$) or $q$ decreases, one can see a significant impact of the parameter on the behaviour of the $q$-deformed oscillator in the phase space. 
First one can see a displacement of the probability function peak towards negative values of the momentum. 
This can be understood by computing the mean values of the $q$-oscillator position and momentum.

The mean value of the oscillator position in stationary states requires the computation of the following integral:
\begin{equation}
\label{x-mean}
\bar x = \int\limits_{ - \infty }^\infty  {\int\limits_{ - \infty }^\infty  
{x\, W_{n,q} \left( {p,x} \right)\rmd x \rmd p} } .
\end{equation}

Using the double sum expression of the Wigner function~(\ref{wq-dsum}), one can see that the integrand is an odd function with respect to~$x$, and therefore~(\ref{x-mean}) is zero.

The mean value of the momentum for stationary states is given by a similar expression:
\begin{equation}
\label{p-mean}
\bar p = \int\limits_{ - \infty }^\infty  {\int\limits_{ - \infty }^\infty  
{p \; W_{n,q} \left( {p,x} \right)\rmd x \rmd p} } .
\end{equation}

Again, one can use the double sum expression of the Wigner function~(\ref{wq-dsum}) to compute this integral, leading to:
\begin{equation}
\label{p}
\bar p= - n \; m \omega h. 
\end{equation}

So the average of the momentum depends on the number of photons and on the finite-difference step~$h$. 
This is the reason for the displacement of the probability function towards negative values of the momentum. 

The last plots in Figs.~1 and~2 show a similar picture. 
In other words the limit $q \rightarrow 0$ dramatically changes the behaviour of the Wigner function. 
The probability distribution function has apparently the form of a Gaussian, and its value does not seem to depend on the photon number~$n$, only its displacement in phase space depends on~$n$. 
This can be understood analytically by computing the limit $h\rightarrow +\infty$ (corresponding to $q\rightarrow 0$). 
In fact, for $n>0$ and for fixed finite values of $x$ and $p$ one finds
\begin{equation}
\lim_{h\rightarrow +\infty} W_{n,q}(p,x) = 0;
\end{equation}
on the other hand, for every $q$ (or $h$):
\begin{equation}
\int\limits_{ - \infty }^\infty  {\int\limits_{ - \infty }^\infty  
W_{n,q} (p,x)\rmd x \rmd p} =1.
\end{equation}
These relations indicate that $\lim_{h\rightarrow +\infty} W_{n,q}(p,x)$, for $n>0$, actually behaves as 
Gaussian distribution function with the peak displaced towards $(-\infty,0)$ in the $(p,x)$-plane.

\begin{figure}
\begin{center}
\begin{tabular}{cc}
\includegraphics[width=0.45\textwidth]{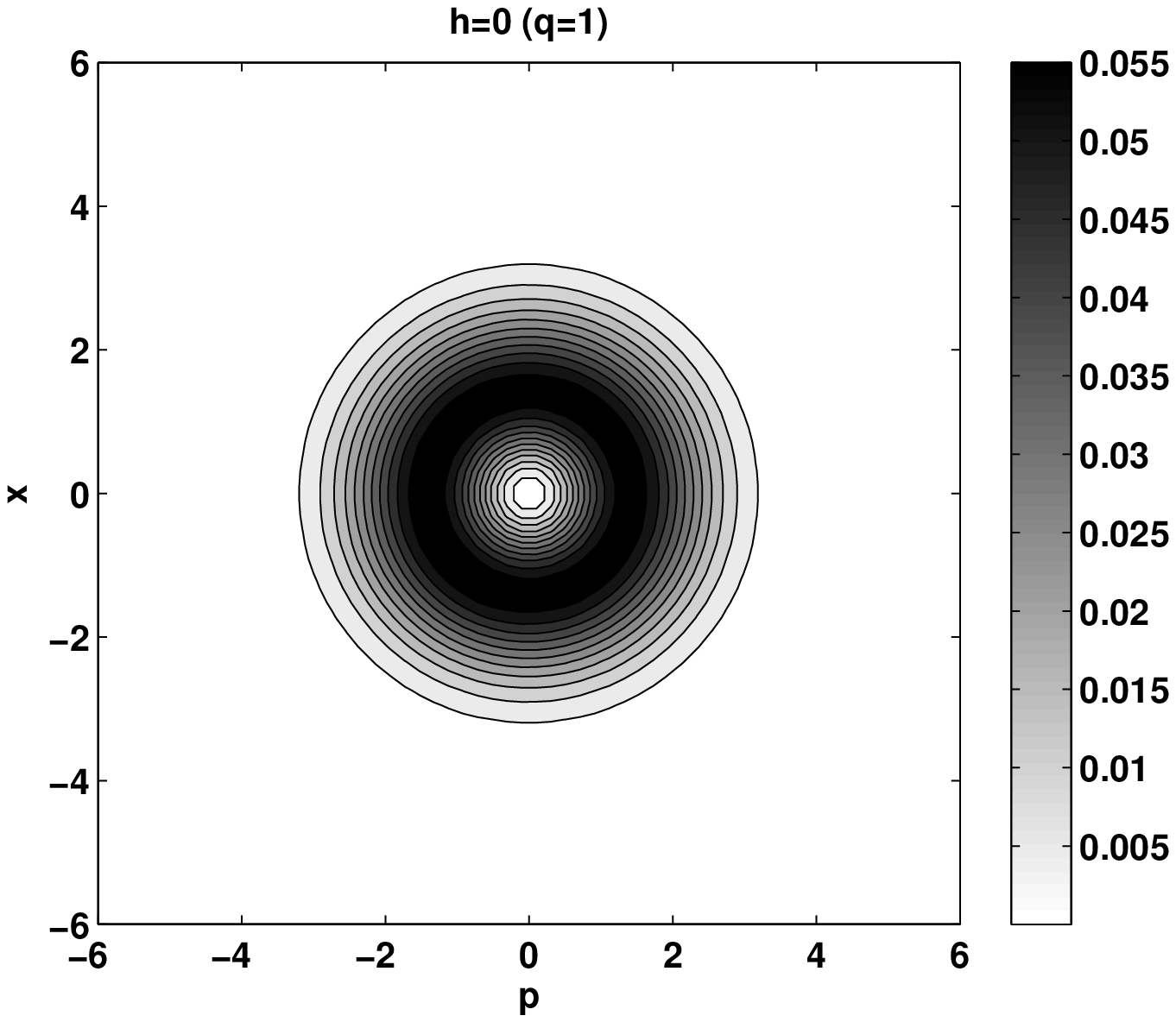}&
\includegraphics[width=0.45\textwidth]{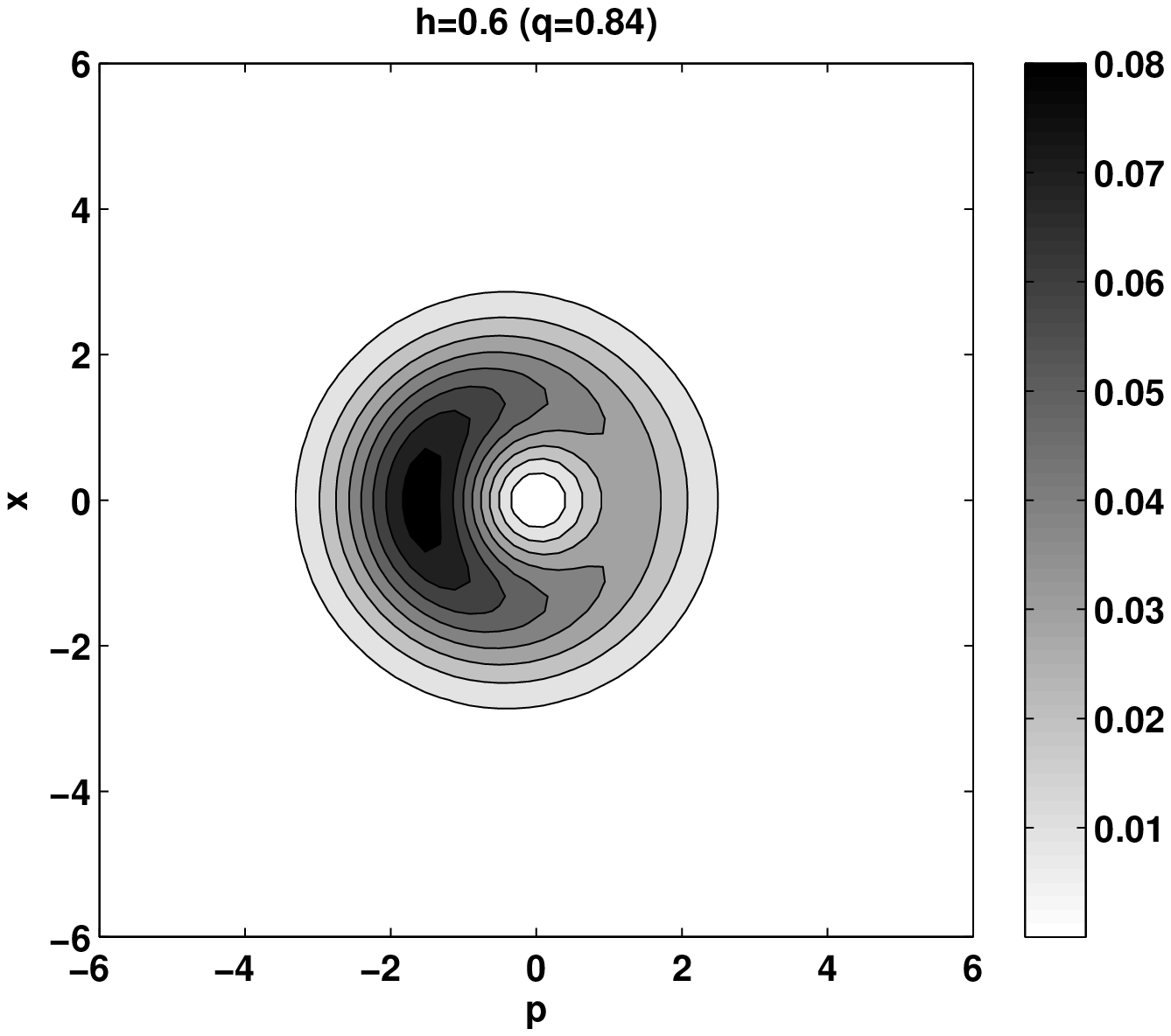}\\
\includegraphics[width=0.45\textwidth]{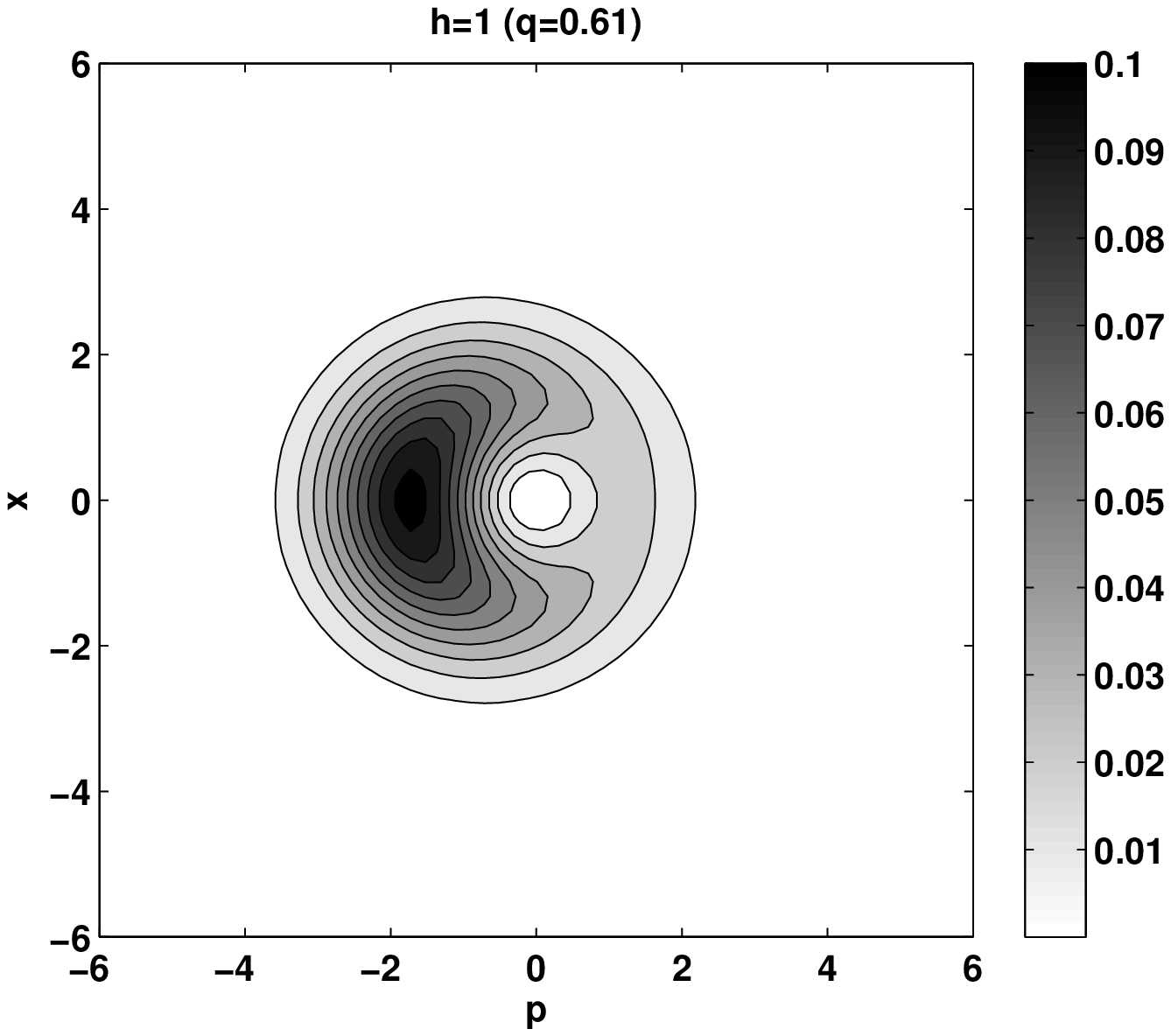}&
\includegraphics[width=0.45\textwidth]{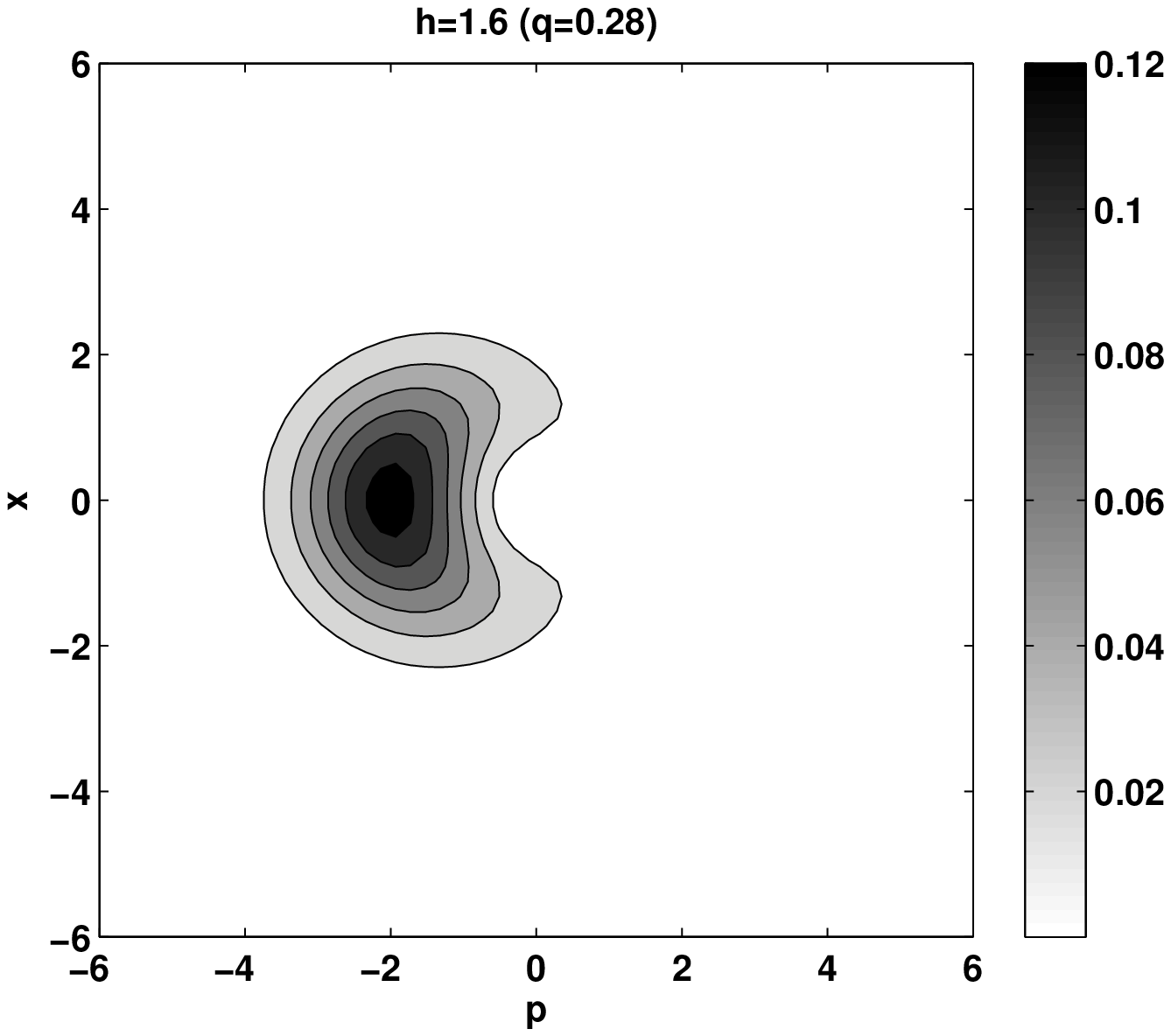}\\
\includegraphics[width=0.45\textwidth]{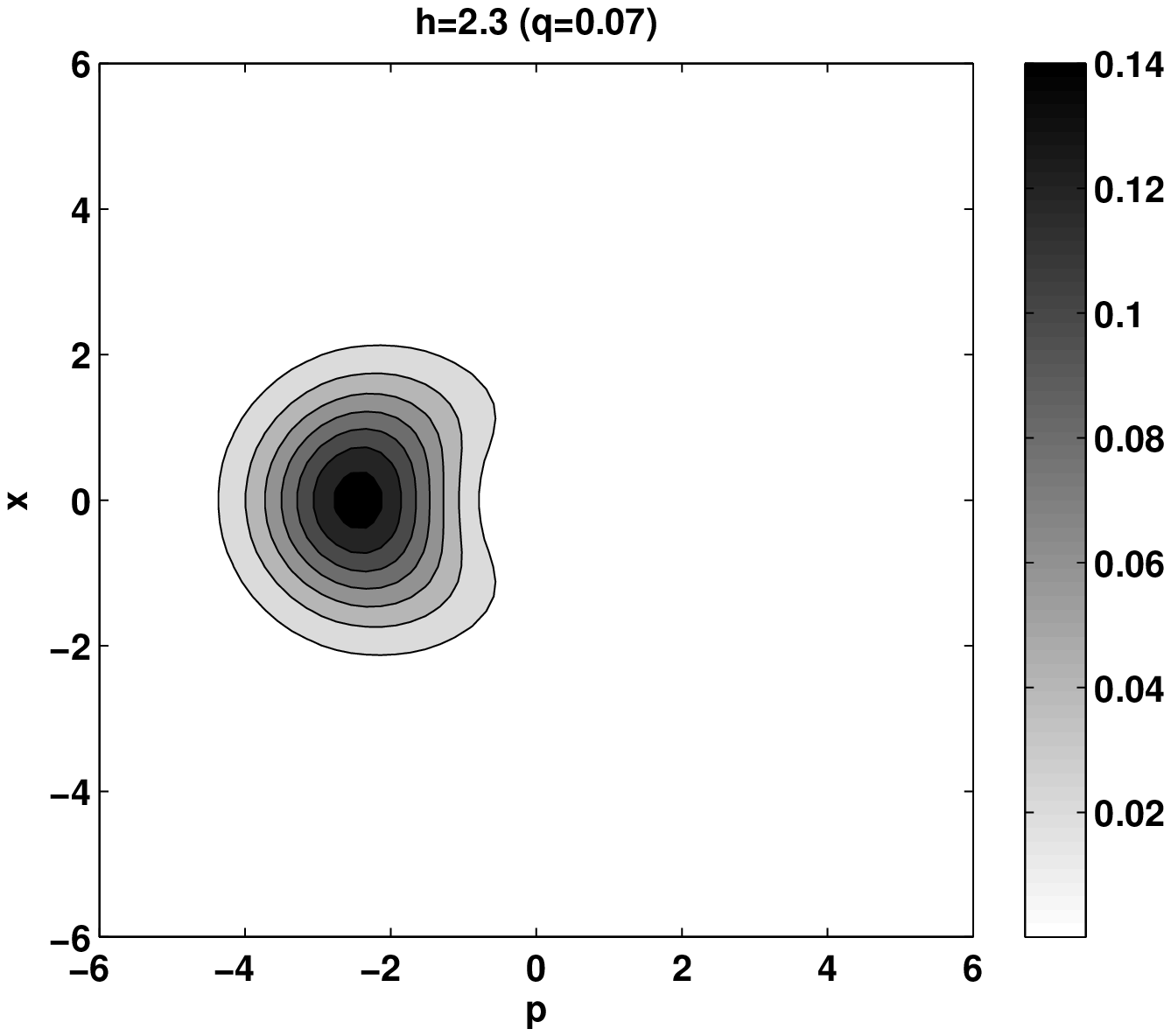}&
\includegraphics[width=0.45\textwidth]{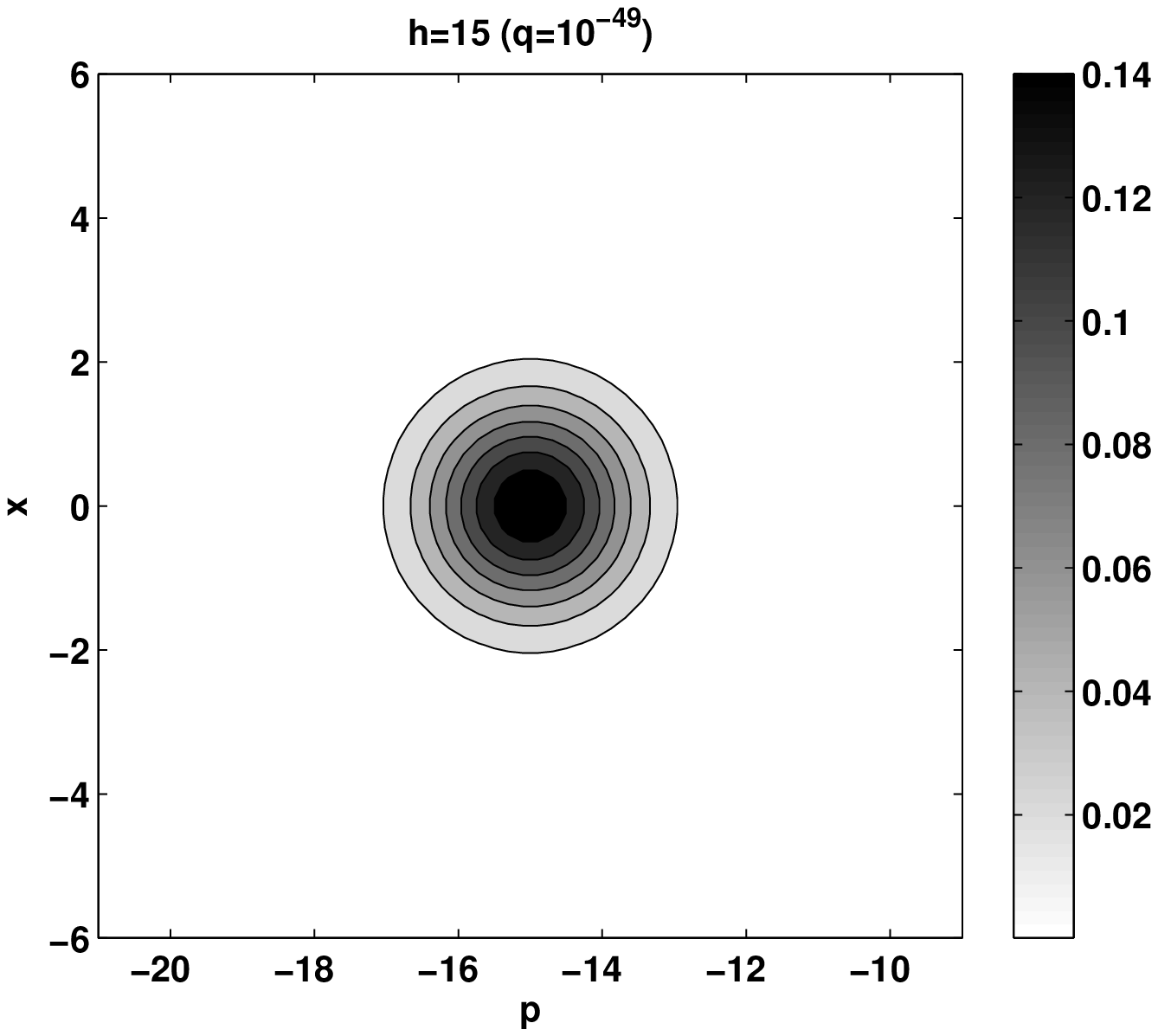}
\end{tabular}
\end{center}
\caption{A density plot of the Husimi function of the single photon state for the $q$-deformed harmonic oscillator, for values of $h=0;\;0.6;\;1;\;1.6;\;2.3;\;15$ ($q=1;\;0.84;\;0.61;\;0.28;\;0.07;\;10^{-49}$) and $m=\omega=\hbar=1$. The behaviour is similar to that of the Wigner function.}
\end{figure}
\begin{figure}
\begin{center}
\begin{tabular}{cc}
\includegraphics[width=0.45\textwidth]{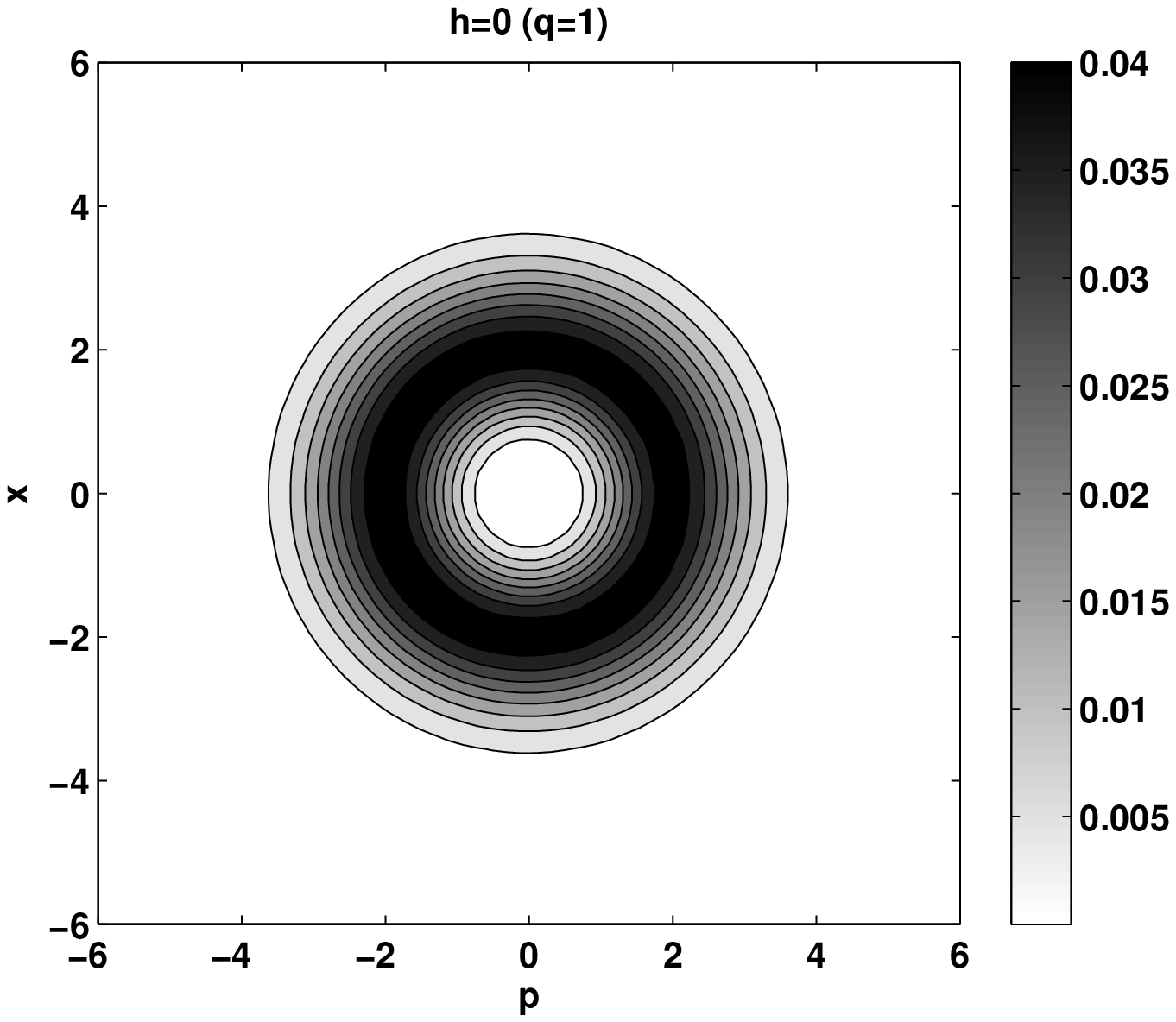}&
\includegraphics[width=0.45\textwidth]{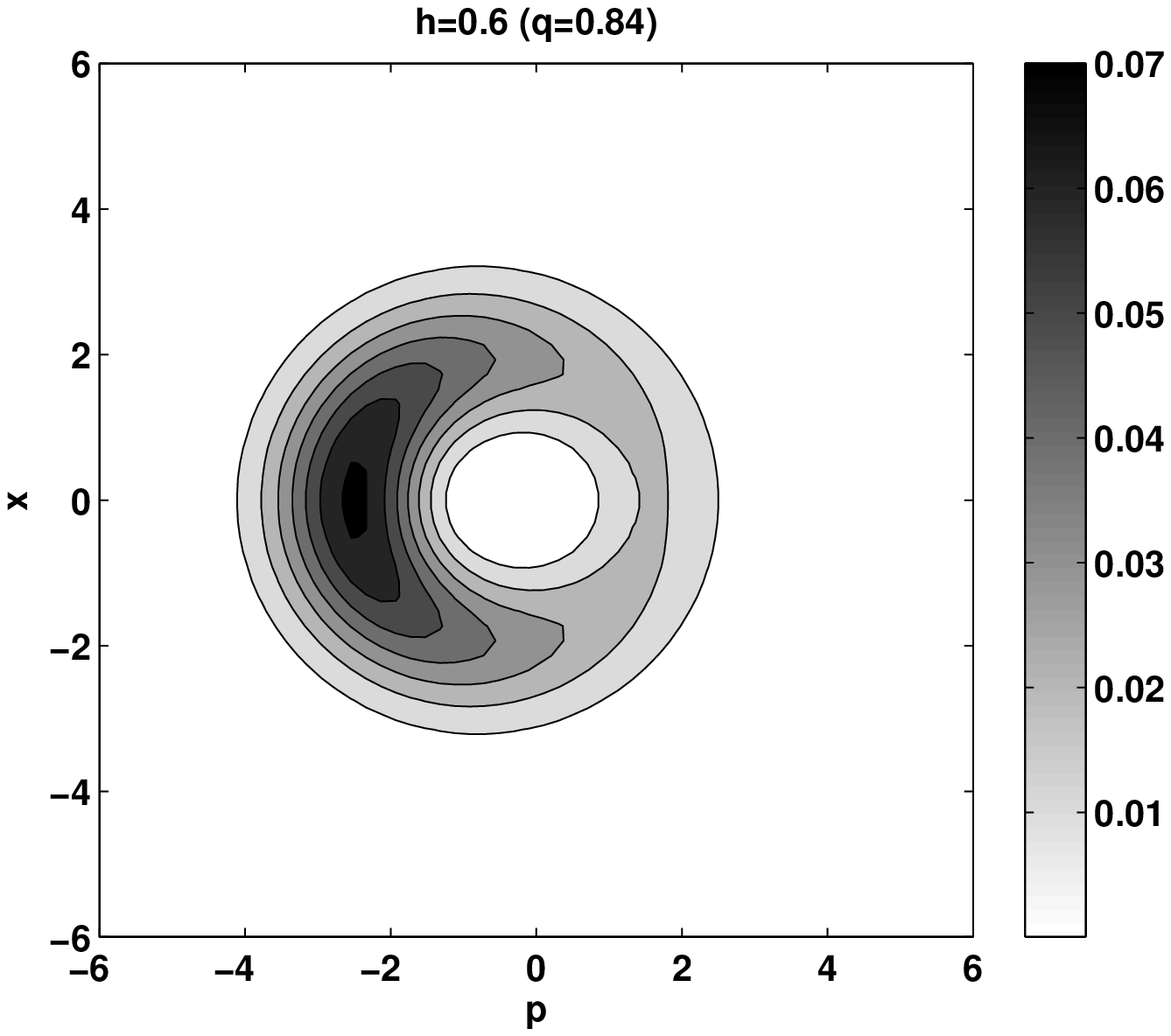}\\
\includegraphics[width=0.45\textwidth]{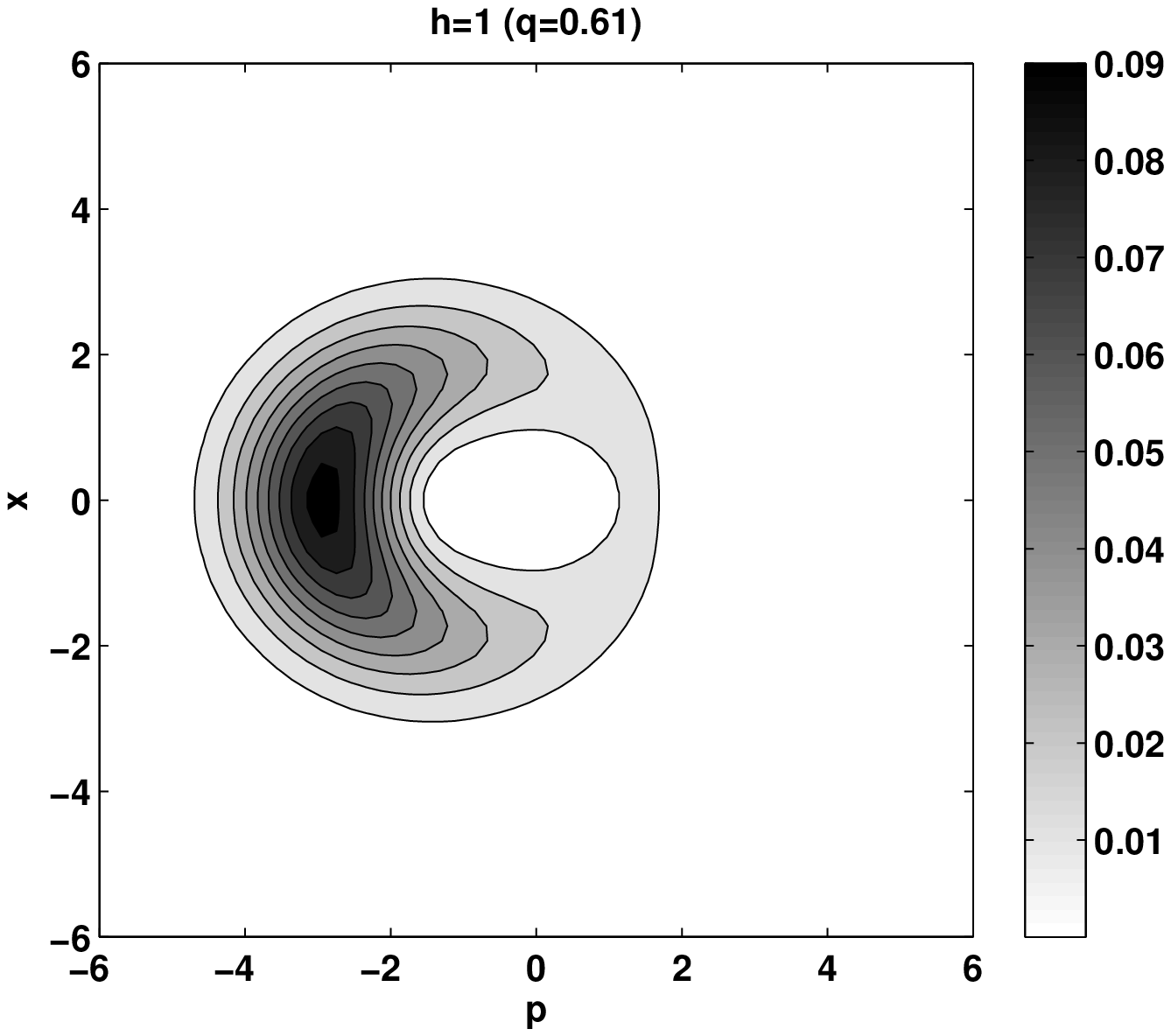}&
\includegraphics[width=0.45\textwidth]{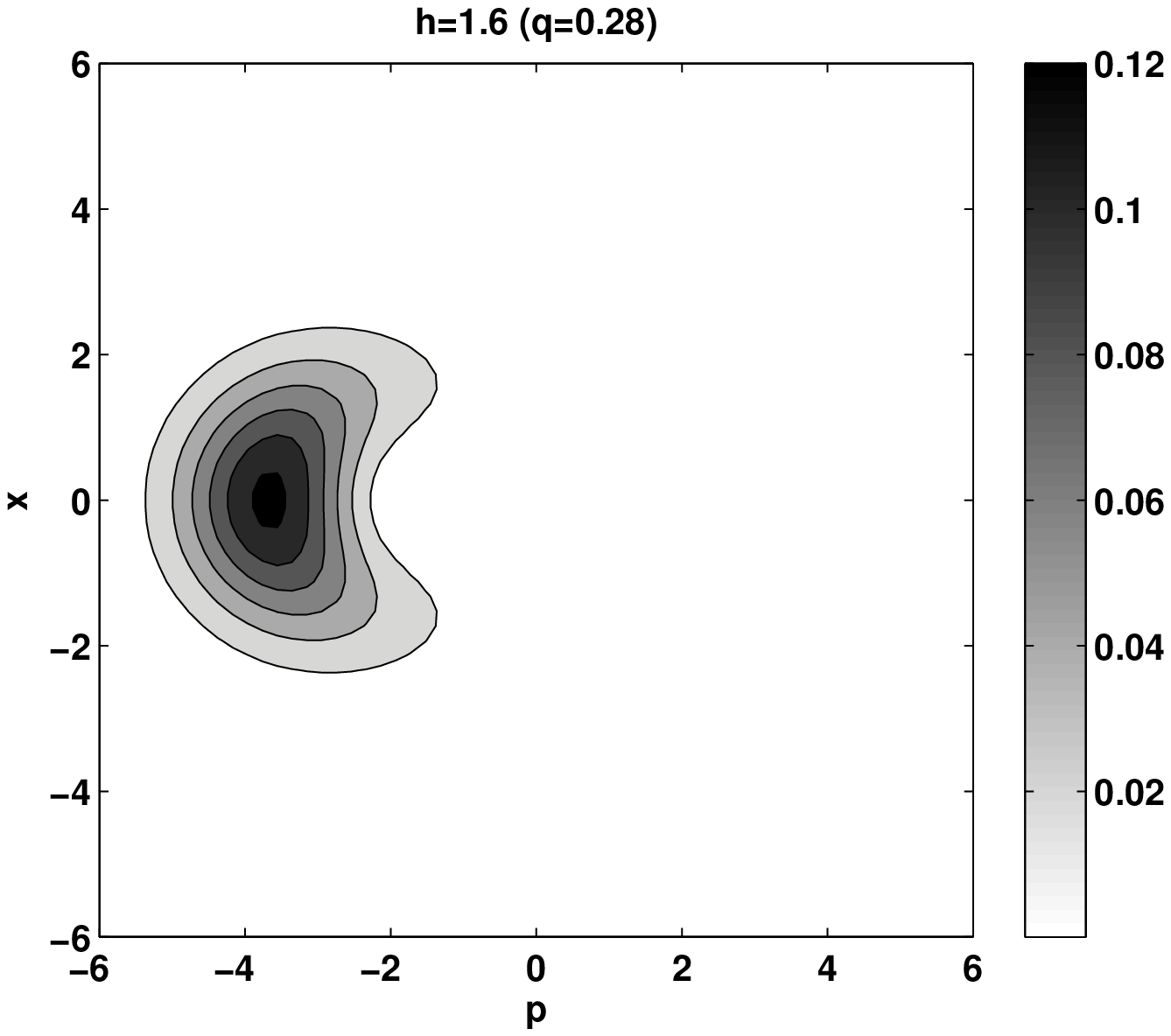}\\
\includegraphics[width=0.45\textwidth]{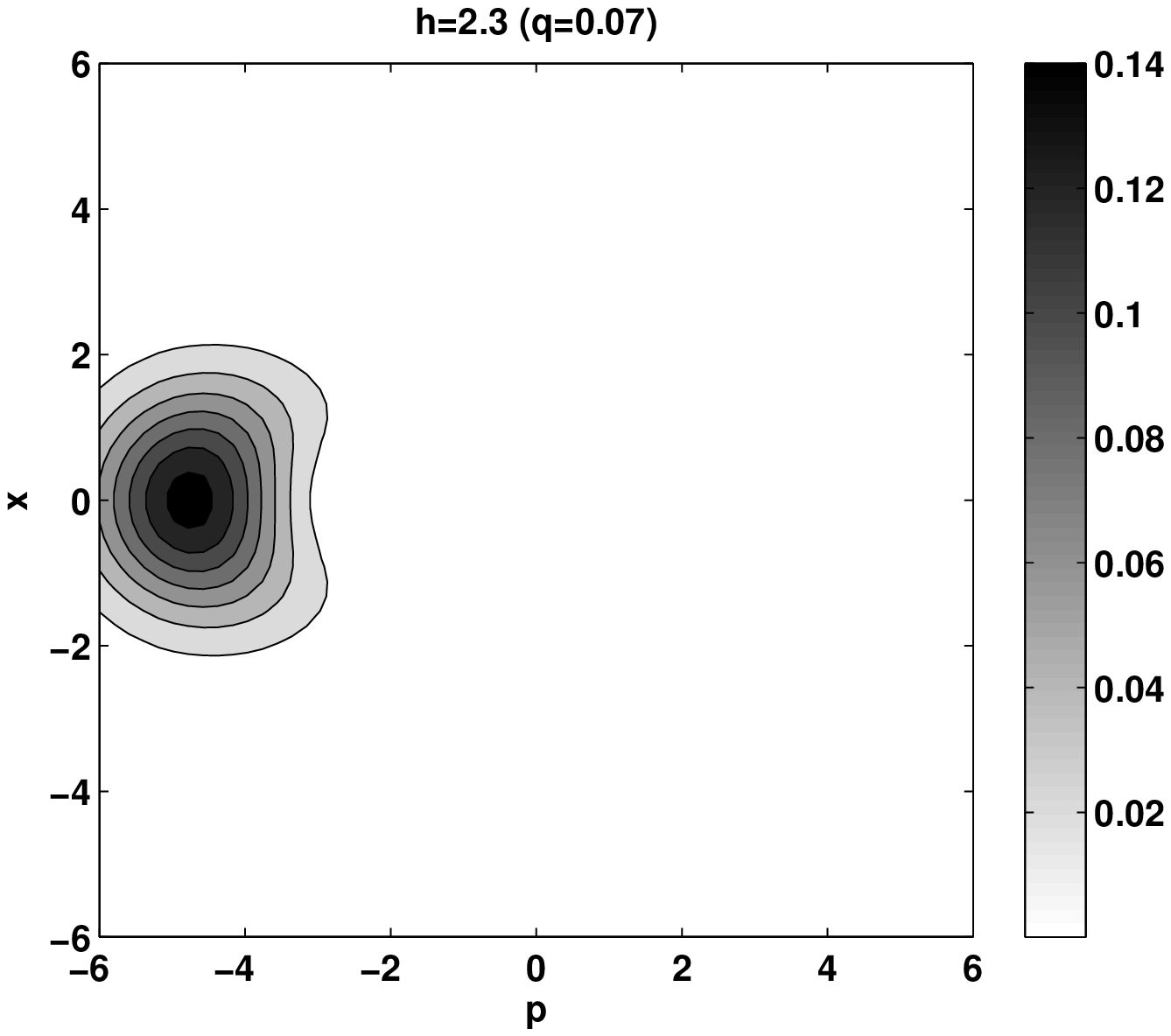}&
\includegraphics[width=0.45\textwidth]{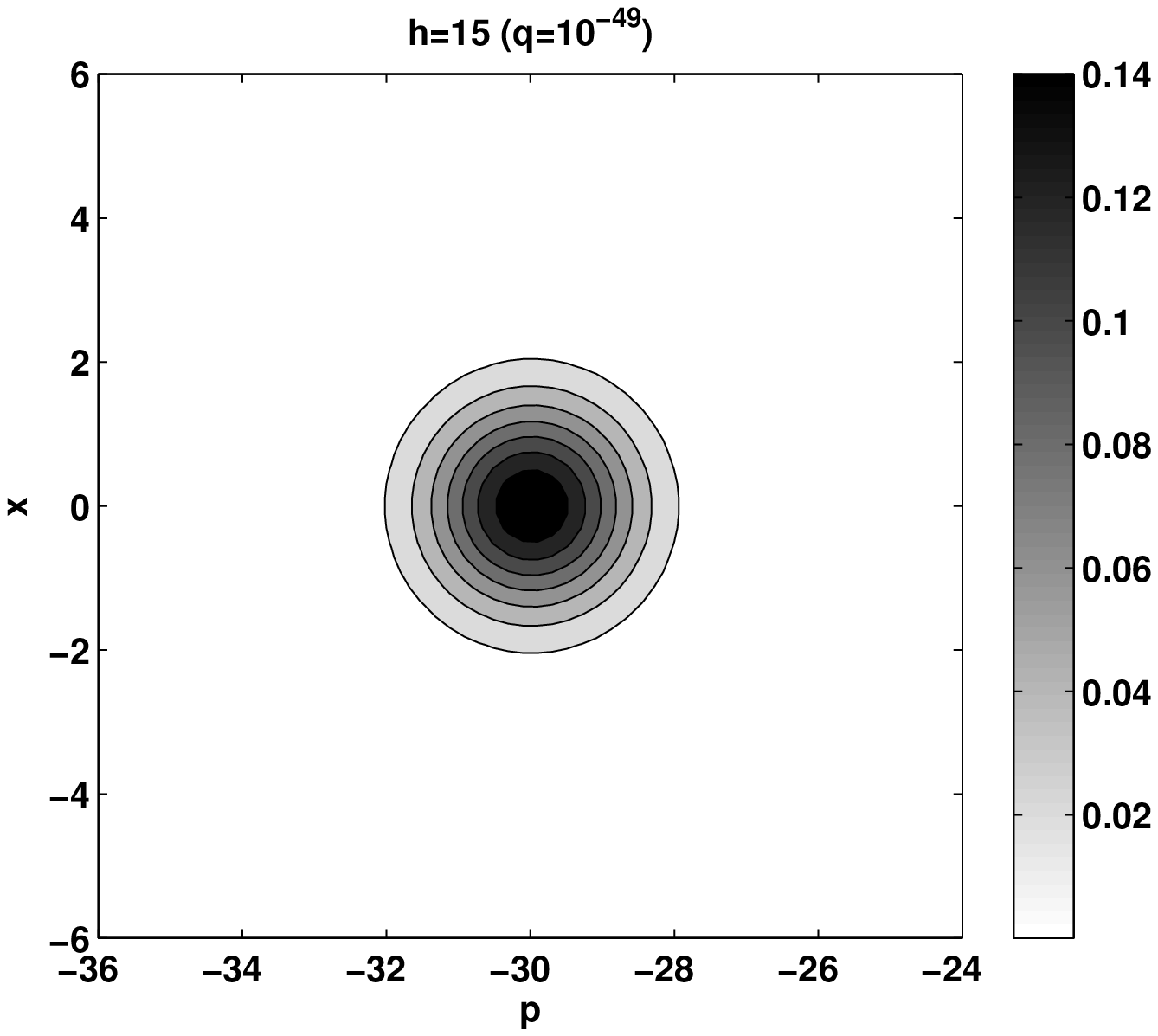}
\end{tabular}
\end{center}
\caption{A density plot of the Husimi function of the double photon state for the $q$-deformed harmonic oscillator, for values of $h=0;\;0.6;\;1;\;1.6;\;2.3;\;15$ ($q=1;\;0.84;\;0.61;\;0.28;\;0.07;\;10^{-49}$) and $m=\omega=\hbar=1$.}
\end{figure}

In Figs.~3 and~4, we also present plots of the Husimi distribution function for the $q$-deformed harmonic oscillator, for $n=1$ and $n=2$. 
The behaviour of the plots in this case is the same as for the Wigner distribution functions. 
Comparing the behaviour of the Wigner and Husimi distributions for small $q$-values ($0<q \ll 1$) indicates that both of them are good approximations of a displaced vacuum state, but the Wigner function is defined more sharply than the Husimi function, allowing a more correct determination of the phase space probabilities of the quantum system under consideration.
In fact, for the Husimi function one can verify that for large $h$-values, \eref{hus-f} is approximated by
\[
\overline W_{n,q}(p,x) \approx \frac{1}{2\pi\hbar} 
\mexp{ - \frac{(p+nm\hbar\omega)^2}{2m\hbar \omega }  -  \frac{m\omega x^2}{2\hbar} }.
\]
In a similar way, one can compute an approximation of the Wigner distribution function for large
$h$-values from~\eref{3phi2}:
\[
W_{n,q}(p,x) \approx \frac{1}{\pi\hbar} 
\mexp{ - \frac{(p+nm\hbar\omega)^2}{m\hbar \omega }  -  \frac{m\omega x^2}{\hbar} }.
\]
These expressions confirm the observed behaviour of the distribution functions as a displaced Gaussian 
when $h$ is large (or when $q$ is close to~0).

\section{Conclusions}

In recent years one has seen an increased interest in the study of quantum systems in phase space, both analytically and experimentally. 
For example, nowadays the development of measurement technologies allows the construction of experimental setups to recover the Wigner distribution function of certain Fock states.
The use of analytical expressions for the Wigner function allows for a more detailed study of quantum entanglement.

We have constructed the Wigner and Husimi distribution functions for the $q$-deformed linear harmonic oscillator, whose wave functions of the stationary states are expressed by means of Rogers-Szeg\"o and Stieltjes-Wigert polynomials in the position and momentum representations. 
The analytical expression of the Wigner distribution function for the stationary states can formally be expressed by means of an Al-Salam-Chihara polynomial.
The Husimi distribution function for the stationary states is simpler and is expressed by means of $q$-shifted factorials. 
Using proper limits for basic hypergeometric series we have shown that the Wigner and Husimi functions of the stationary states reduce to those for the non-relativistic quantum harmonic oscillator, when $q\rightarrow 1$.

Examination of both distribution functions in the $q$-model shows that, when $q$ tends to $0$, their behaviour in phase space is similar to the ground state of an ordinary quantum oscillator, but with a displacement of momentum towards negative values. 
We have computed the mean values of the position and momentum for the Wigner function and, unlike the ordinary case, the mean values of the momentum are not zero but depend on $q$ and $n$. 
The ground-state like behaviour of the distribution functions for excited states in the $q$-case opens up new perspectives for further experimental measurements of quantum systems in the phase space.

\ack

One of the authors (E.I.J.) would like to acknowledge that this work is performed in the framework of the Fellowship 05-113-5406 under the INTAS-Azerbaijan YS Collaborative Call 2005.


\section*{References}

\begin{thebibliography}{99}

\bibitem{moshinsky}
Moshinsky M 1969 {\it The Harmonic Oscillator in Modern Physics: from Atoms to Quarks} (New-York: Gordon and Breach)

\bibitem{landau}
Landau L D and  Lifshitz E M 1997 {\it Quantum Mechanics: Non-Relativistic Theory} (Oxford: Butterworth-Heinemann)

\bibitem{wigner}
Wigner E P 1932 {\PR} {\bf 40} 749

\bibitem{smithey}
Smithey D T, Beck M, Raymer M G and Faridani A 1993 \PRL {\bf 70} 1244

\bibitem{mcmahon}
McMahon P J, Allman B E, Jacobson D L, Arif M, Werner S A and Nugent K A 2003 \PRL {\bf 91} 145502

\bibitem{lvovsky}
Lvovsky A I, Hansen H, Aichele T, Benson O, Mlynek J and Schiller S 2001 \PRL {\bf 87} 050402

\bibitem{ourjoumtsev}
Ourjoumtsev A, Tualle-Brouri R and Grangier P 2006 \PRL {\bf 96} 213601

\bibitem{leibfried}
Leibfried D, Meekhof D M, King B E, Monroe C, Itano W M and Wineland D J 1996 \PRL {\bf 77} 4281

\bibitem{iwata}
Iwata G 1951 {\it Prog. Theor. Phys.} {\bf 6} 524

\bibitem{greenberg}
Greenberg O W 1990 \PRL {\bf 64} 705; 1991 \PR {\bf D43} 4111

\bibitem{chaturvedi}
Chaturvedi S, Kapoor A K, Sandhya R, Srinivasan V and Simon R 1991 \PR {\bf A43} 4555

\bibitem{lavagno}
Lavagno A, Scarfone A M and Narayana Swamy P 2005 {\it Rep. Math. Phys.} {\bf 55} 423

\bibitem{ignatiev}
Ignatiev A Yu 2006 {\it Rad. Phys. Chem.} {\bf 75} 2090

\bibitem{rajagopal}
Rajagopal A K 1993 \PR {\bf A47} R3465

\bibitem{zhang}
Zhang Q H and Padula S S 2004 \PR {\bf C69} 024907

\bibitem{galetti}
Galetti D, Mizrahi S S and Ruzzi M 2004 \JPA {\bf 37} L643

\bibitem{GR2}
Gasper G and Rahman M 2004 {\it Basic hypergeometric series, Encyclopedia of Mathematics And Its Applications 96} (Cambridge: Cambridge University Press)

\bibitem{hillery}
Hillery M, O'Connell R F, Scully M O and Wigner E P 1984 {\it Phys. Rep.} {\bf 106} 121

\bibitem{tatarskii}
Tatarskii V I 1983 {\it Sov. Phys. Uspekhi} {\bf 26} 311

\bibitem{davies}
Davies R W and Davies K T R 1975 \APNY {\bf 89} 261

\bibitem{KoeSwart}
Koekoek R and Swarttouw R F 1998 {\it The Askey-scheme of hypergeometric orthogonal polynomials and its q-analogue} (Delft University of Technology: Report no. 98-17)

\bibitem{royer}
Royer A 1977 \PR {\bf A15} 449

\bibitem{cartwright}
Cartwright N D 1975 {\it Physica} {\bf A83} 210

\bibitem{janssen}
Janssen A J E M 1981 {\it SIAM J. Math. Anal.} {\bf 12} 752

\bibitem{rajagopal1}
Rajagopal A K 1983 \PR {\bf A27} 558

\bibitem{husimi}
Husimi K 1940 {\it Proc. Phys. Math. Soc. Jpn.} {\bf 22} 264

\bibitem{macfarlane}
Macfarlane A J 1989 \JPA {\bf 22} 4581

\bibitem{kagramanov}
Kagramanov E D, Mir-Kasimov R M and Nagiyev S M 1990 \JMP {\bf 31} 1733

\bibitem{atakishiyev}
Atakishiyev N M and Suslov S K 1990 {\it Theor. Math. Phys.} {\bf 85} 1055; 1991 {\it Theor. Math. Phys.} {\bf 87} 442

\bibitem{mirkasimov}
Mir-Kasimov R M 1991 \JPA {\bf 24} 4283

\bibitem{vanderjeugt}
Van der Jeugt J 1992 {\it Lett. Math. Phys.} {\bf 24} 267; 1993 \JMP {\bf 34} 1799

\bibitem{atakishiyev2}
Atakishiyev N M, Frank A and Wolf K B 1994 \JMP {\bf 35} 3253

\bibitem{atakishiyev3}
Atakishiyev N M and Nagiyev S M 1994 {\it Theor. Math. Phys.} {\bf 98} 162

\bibitem{nagiyev}
Nagiyev S M 1995 {\it Theor. Math. Phys.} {\bf 102} 180

\bibitem{atakishiyev4}
Atakishiyev M N, Atakishiyev N M and Klimyk A U 2006 \JMP {\bf 47} 093502

\bibitem{atakishiyev5}
Atakishiyev N M and Nagiyev S M 1994 \JPA {\bf 27} L611

\bibitem{Ismail}
Ismail M E H, Dunkl C F and Wong R 2000 {\it Special Functions} (World Scientific) 

\bibitem{wall}
Wall F T 1986 {\it Proc. Natl. Acad. Sci. U.S.A.} {\bf 83} 5360

\bibitem{march}
March N H 2005 {\it Int. J. Quantum Chem.} {\bf 105} 701

\end{thebibliography}
\end{document}